\documentclass[11pt]{article}
\usepackage{times}
\usepackage{graphicx}
\usepackage{authblk}
\usepackage[a4paper,margin=1in]{geometry}
\usepackage{hyperref}
\usepackage{xcolor}
\usepackage{tcolorbox}
\usepackage{amsmath,amssymb}
\usepackage{fancyhdr}
\usepackage{caption}
\usepackage{float}
\usepackage{algorithm}
\usepackage{algorithmic}
\usepackage{subcaption}
\usepackage{caption}
\usepackage{framed}

\hypersetup{
    colorlinks=true,
    linkcolor=blue,
    citecolor=blue,
    urlcolor=blue
}

\makeatletter
\renewcommand{\fnum@algorithm}{\textbf{Algorithm~\thealgorithm}}
\makeatother

\title{\textbf{Explainable Vulnerability Detection in C/C++ Using Edge-Aware Graph Attention Networks}}

\author[1]{Radowanul Haque}
\author[1]{Aftab Ali}
\author[1]{Sally McClean}
\author[1]{Naveed Khan}
\affil[1]{School of Computing, Ulster University, United Kingdom}

\date{}

\begin{document}

\maketitle

\begin{abstract}
Detecting security vulnerabilities in source code remains challenging, particularly due to class imbalance in real-world datasets where vulnerable functions are under-represented. Existing learning-based methods often optimise for recall, leading to high false positive rates and reduced usability in development workflows. Furthermore, many approaches lack explainability, limiting their integration into security workflows. This paper presents ExplainVulD, a graph-based framework for vulnerability detection in C/C++ code. The method constructs Code Property Graphs and represents nodes using dual-channel embeddings that capture both semantic and structural information. These are processed by an edge-aware attention mechanism that incorporates edge-type embeddings to distinguish among program relations. To address class imbalance, the model is trained using class-weighted cross-entropy loss. ExplainVulD achieves a mean accuracy of 88.25\% and an F1 score of 48.23\% across 30 independent runs on the ReVeal dataset. These results represent relative improvements of 4.6\% in accuracy and 16.9\% in F1 score compared to the ReVeal model, a prior learning-based method. The framework also outperforms static analysis tools, with relative gains of 14.0–14.1\% in accuracy and 132.2–201.2\% in F1 score. Beyond improved detection performance, ExplainVulD produces explainable outputs by identifying the most influential code regions within each function, supporting transparency and trust in security triage.

\end{abstract}

\section{Introduction}

Software vulnerabilities in low-level languages such as C and C++ continue to pose significant risks across a range of domains, including browser engines, operating system kernels, embedded systems, and Internet-of-Things (IoT) devices. Exploitable flaws in such systems can lead to remote code execution, data leakage, or full system compromise~\cite{max-mergin}. Traditional approaches such as static analysis and manual code review are widely used in practice to detect these issues. Static analysis tools, including Flawfinder~\cite{wheeler_flawfinder} and Cppcheck~\cite{cppcheck}, operate by scanning source code for syntactic patterns or predefined rules that are indicative of common vulnerabilities~\cite{1quDeASK}. Manual review, in contrast, involves human inspection of code to reason about program logic, control flow, and semantic correctness. While these techniques are valuable, they face significant limitations. Static tools tend to produce a high number of false positives and lack contextual understanding, while manual review is time-consuming, error-prone, and difficult to scale to large codebases~\cite{1quDeASK,whynotstatic}.

In response, learning-based approaches have been proposed as alternatives for detecting vulnerabilities in source code. These methods aim to automatically identify patterns associated with known vulnerabilities, reducing dependence on manually crafted rules. Earlier models include VulDeePecker~\cite{zou2019uvuldeepecker}, $\mu$VulDeePecker~\cite{zou2019mu}, SySeVR~\cite{9321538-sysevr}, and VulDeeLocator~\cite{vuldelocator}, while graph-based approaches include Devign~\cite{zhou2019devign}, BGNN4VD~\cite{BGNN4VD}, ReVeal~\cite{chakraborty2021deep}, and IvDetect~\cite{1ivdetc}.

\begin{figure}[!t]
\centering
\includegraphics[width=0.35\linewidth]{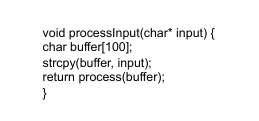}
\caption{Example C function with a buffer overflow vulnerability.}
\label{fig:example_vuln}
\end{figure}

Sequence-based models, such as VulDeePecker and SySeVR, treat source code as a linear token sequence, following methods from natural language processing. These models can learn lexical patterns but do not capture control or data dependencies between code elements~\cite{chakraborty2021deep}. Their performance is often sensitive to small changes in structure. In contrast, graph-based models represent source code as structured graphs, such as Abstract Syntax Trees (ASTs), Control Flow Graphs (CFGs), and Data Flow Graphs (DFGs). These representations enable models to reason over both syntactic structure and execution flow~\cite{cpg}.
To illustrate the challenge of vulnerability detection, consider the C function from Figure~\ref{fig:example_vuln}.

This function contains a classic buffer overflow vulnerability. The call to \texttt{strcpy} copies the contents of \texttt{input} into a fixed-size buffer without verifying the input length. If the input exceeds the allocated space, it can overwrite adjacent memory, potentially altering program state or control flow. While this example appears straightforward, similar vulnerabilities often occur in more complex forms across large codebases, making them difficult to detect using static pattern matching alone.

A Code Property Graph (CPG)~\cite{cpg} provides a unified representation by combining Abstract Syntax Trees (ASTs), Control Flow Graphs (CFGs), and Data Flow Graphs (DFGs). For the above function, the CPG encodes the buffer declaration, the data flow from \texttt{input} to \texttt{strcpy}, and the use of \texttt{buffer} in the return statement. These relationships are captured as typed edges, enabling graph-based models to reason about how information propagates through the code in terms of both control and data dependencies.

Despite this progress, most existing graph-based methods suffer from important limitations. They often use flat or shallow node features derived from syntax or tokens, and typically ignore the semantics of edge types that define relationships within the graph. Furthermore, the majority of these models act as black boxes, offering limited insight into why a function is predicted to be vulnerable. In software security workflows, understanding the reasoning behind a prediction is critical for trust, verification, and remediation.

In this work, we introduce ExplainVulD, a graph-based framework for explainable vulnerability detection in C/C++ functions. The framework represents each function as a Code Property Graph (CPG) constructed using Joern, where nodes correspond to code elements such as identifiers and control structures, and edges capture structural and semantic relationships. We construct dual-channel node embeddings to capture both semantic and structural aspects of code. The semantic channel uses a Word2Vec~\cite{mikolov2013efficientestimationwordrepresentations} model trained on token sequences derived from filtered AST node labels, capturing lexical usage patterns. The structural channel is built from a second Word2Vec model trained on metapath-guided random walks over the CPG, incorporating node types, edge types, AST depth differences, and semantic scopes such as return paths. The two channels are concatenated to form a 1024-dimensional node embedding.

To propagate information, we apply GATv2~\cite{gatv2} extended with an edge-aware attention mechanism that incorporates learned edge-type embeddings. This enables the network to differentiate between relation types during message passing, supporting reasoning over heterogeneous program structures.

ExplainVulD also includes a post hoc explanation module. We compute relevance scores for nodes and edges using a combination of attention weights from global pooling and input gradients with respect to the classification loss. The most influential elements are then mapped back to source-level constructs, allowing structured interpretation of the model’s decisions. In contrast to general-purpose tools such as GNNExplainer~\cite{ying2019gnnexplainergeneratingexplanationsgraph}, our approach is model-specific and operates directly on the CPG representation.

By jointly modelling semantic and structural properties, incorporating relation-sensitive message passing, and producing interpretable outputs, ExplainVulD addresses common limitations of prior graph-based approaches to vulnerability detection.

\subsection{Contributions}

This work makes the following contributions to graph-based vulnerability detection in C/C++ code:

\begin{itemize}
    \item To the best of our knowledge, this is the first work to propose a dual-channel node embedding for CPGs that combines semantic representations from AST tokens with structural features derived from metapath-guided walks.

    \item We extend GATv2 to incorporate edge-type embeddings into the attention mechanism, enabling relation-sensitive message passing over heterogeneous code graphs.
    
    \item We design an explanation module that combines attention weights and input gradients to identify influential nodes and edges in the CPG. While similar techniques exist in other domains, to our knowledge this is the first application of such a method to graph-based vulnerability detection in code.

    \item We evaluate ExplainVulD on the ReVeal dataset using 30 independent runs with randomised splits. Our experiments compare the method against static and learning-based baselines and include ablation and explainability analyses.
\end{itemize}

The remainder of the paper is structured as follows. Section~\ref{sec:related_work} reviews related work on vulnerability detection and graph learning. Section~\ref{sec:overview} outlines the overall framework. Section~\ref{sec:methodology} describes the embedding design and model architecture. Section~\ref{sec:experimental_setup} details the experimental setup. Section~\ref{sec:results} presents the results and analysis. Section~\ref{discussion} discusses explainability and ablation findings. Finally, Section~\ref{conclusion} concludes the paper.

\section{Related Works}
\label{sec:related_work}

Early methods for software vulnerability detection primarily relied on static analysis tools. These tools, such as \textit{Flawfinder}~\cite{wheeler_flawfinder}, identify vulnerabilities by matching source code against pre-defined patterns and rules~\cite{KAUR20202023-4}. Static analysers can effectively detect simple and common vulnerabilities; however, they often produce many false positives. Additionally, these methods struggle with vulnerabilities that require deeper semantic or contextual understanding of code behaviour~\cite{KAUR20202023-4}, \cite{russell2018automated}.

To address these limitations, researchers explored deep learning-based approaches, starting with sequence-based models. These methods treat source code as sequences of tokens, similar to how natural language models handle text. For instance, VulDeePecker~\cite{zou2019uvuldeepecker} uses an LSTM neural network to identify vulnerabilities related to buffer overflows and improper resource management. Further improvements include contextual LSTM methods~\cite{8599360-19}, which use neighbouring code snippets to better capture context, and KELM~\cite{https://doi.org/10.1155/2021/5566423-kelm}, which combines doc2vec embeddings with kernel-based learning for faster and more accurate detection. Recent work by Shestov et al.\ demonstrates that finetuning large pretrained models like WizardCoder improves vulnerability detection performance in Java code, especially under class imbalance~\cite{fine-tune_llm}. However, sequence-based methods lack an understanding of structural relationships in code, such as data flow or control flow dependencies, limiting their effectiveness for certain complex vulnerabilities~\cite{TANG2023111623-csgvd}, \cite{10.1145/3510454.3516865-regvd}.

Graph-based methods emerged to overcome these structural limitations. These techniques represent code as graphs, typically using Abstract Syntax Trees (ASTs), Control Flow Graphs (CFGs), Data Flow Graphs (DFGs), or a unified representation known as a Code Property Graph (CPG) ~\cite{cpg}. SySeVR~\cite{9321538-sysevr} is an example that utilises multiple graph representations to enhance detection accuracy, while ReGVD~\cite{10.1145/3510454.3516865-regvd} further refines graph construction methods to improve generalisation. ReVeal~\cite{chakraborty2021deep} applies Graph Neural Networks (GNNs) directly to CPGs, achieving strong results on imbalanced, real-world datasets. Nevertheless, graph-based models often lack interpretability, meaning they do not clearly indicate which parts of the code contribute to their predictions~\cite{TANG2023111623-csgvd}, \cite{10.1145/3510454.3516865-regvd}.

Hybrid methods combine features from both sequence and graph representations to capture broader aspects of source code. Models such as CSGVD~\cite{TANG2023111623-csgvd} and VUDENC~\cite{WARTSCHINSKI2022106809-vudenc} integrate token-level embeddings (e.g., Word2Vec~\cite{mikolov2013efficientestimationwordrepresentations}, Code2Vec~\cite{code2vec}) with graph-based structural features. These hybrid approaches show improved accuracy in detecting complex vulnerabilities involving combined structural and token-level patterns. However, they still tend to operate as black boxes, offering limited insight into the reasoning behind their decisions.

In response to these issues, our proposed approach introduces an interpretable and edge-aware framework for within-project vulnerability detection. We use dual-channel embeddings, integrating semantic and structural information explicitly, and an edge-aware Graph Attention Network (GATv2) ~\cite{gatv2} architecture to model relationships between nodes. Our method not only improves detection accuracy but also clearly identifies the subgraphs contributing to each prediction, providing interpretability. To our knowledge, this is the first framework to apply edge-aware attention mechanisms to CPG-based vulnerability detection with explicit interpretability.

\section{Overview of the ExplainVulD Framework}
\label{sec:overview}

ExplainVulD is a graph-based framework for detecting vulnerabilities in C/C++ functions by learning from both semantic and structural aspects of source code. It operates on Code Property Graphs (CPGs), which integrate multiple representations of a program into a unified graph structure. The framework comprises four core components: CPG construction, dual-channel node embedding, edge-aware message passing using graph attention, and explanation generation.

Each function is first converted into a CPG using Joern, which combines Abstract Syntax Trees (ASTs), Control Flow Graphs (CFGs), and Data Flow Graphs (DFGs). Nodes represent program elements such as identifiers, expressions, or control statements, while edges encode syntactic, control, and data dependencies.

Node features are constructed using a dual-channel embedding scheme. The semantic channel encodes lexical information using a Word2Vec~\cite{mikolov2013efficientestimationwordrepresentations} model trained on sequences of code tokens extracted from filtered AST node labels. The structural channel captures context through metapath-guided random walks, which incorporate node types, edge types, and traversal patterns. These walks reflect relationships such as dominance, reachability, and use-definition chains, and are used to train a second Word2Vec model. The semantic and structural embeddings are concatenated to produce the final node representation.

The resulting graph is processed using two GATv2~\cite{gatv2} layers, modified to incorporate edge-type embeddings directly into the attention mechanism. This design enables relation-specific message passing by allowing the network to distinguish between different edge types during information propagation. A global attention pooling layer aggregates node-level representations into a graph-level embedding, which is passed to a multilayer perceptron (MLP)~\cite{MLP} for binary classification.

To support interpretability, ExplainVulD includes a dedicated explanation module. It computes relevance scores for nodes and edges by combining attention weights with input gradients from the classification loss. The top-ranked components are then mapped back onto the source code via the CPG, allowing identification of the code regions most influential to the model’s prediction.

\begin{figure*}[!t]
\centering
\includegraphics[width=\textwidth]{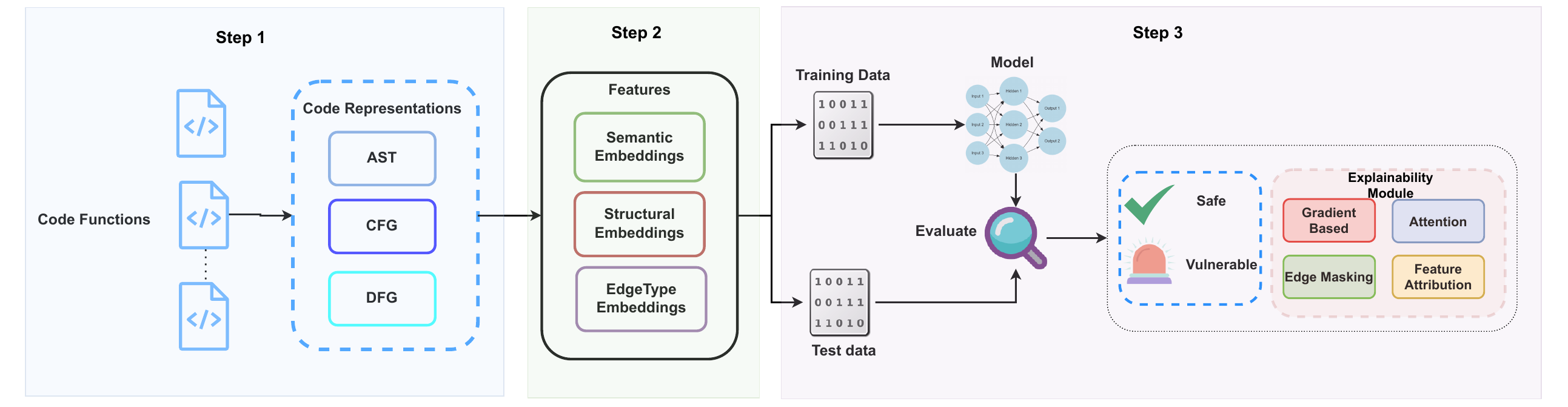}
\caption{Overview of the ExplainVulD framework.}
\label{fig_1}
\end{figure*}

A schematic overview of the full framework is shown in Figure~\ref{fig_1}.

\section{Methodology}
\label{sec:methodology}

This section outlines the design of \textit{ExplainVulD}, detailing the data processing pipeline, feature construction, model architecture, and training procedure used for vulnerability detection in C/C++ code.

\subsection{Problem Formulation}

Let \( G = (V, E, T) \) denote a Code Property Graph (CPG), where \( V \) is the set of nodes corresponding to program entities such as expressions, control blocks, and variables. The set \( E \subseteq V \times V \) contains directed edges capturing semantic, syntactic, and control-flow relationships. Each edge is annotated by a type mapping \( T : E \rightarrow \{1, 2, \dots, 14\} \), where each integer encodes a specific relation type (e.g., \texttt{controls}, \texttt{def}, \texttt{flows\_to}, \texttt{is\_ast\_parent}).

Each node \( v_i \in V \) is associated with a feature vector \( \mathbf{x}_i \in \mathbb{R}^d \) that encodes both structural and semantic characteristics. Every edge \( (v_i, v_j) \in E \) is labelled with a categorical type \( t_{ij} = T(i, j) \in \{1, \dots, 14\} \), which is later embedded into a continuous representation for use during edge-aware message passing.

Given a dataset of graph-labelled samples \( \{(G_k, y_k)\}_{k=1}^N \), where \( y_k \in \{0, 1\} \) denotes whether function \( k \) is vulnerable, the objective is to learn a graph classification function
\begin{equation}
f(G; \theta) : G \rightarrow \hat{y} \in [0, 1]^2,
\label{eq:graph_classification}
\end{equation}
parameterised by \( \theta \), which outputs a predicted probability distribution over the binary classes.

In practice, vulnerable functions constitute only a small fraction of the dataset. This class imbalance presents two major challenges: (i) the need to identify subtle patterns that characterise vulnerable code, and (ii) the risk of model bias towards the dominant class during training.

To address these challenges, we propose a framework that incorporates: (i) multi-view node representations combining semantic and structural features; (ii) edge-aware attention for relation-sensitive aggregation; and (iii) a class-weighted cross-entropy loss function to emphasise hard and minority-class examples. The following subsections detail each of these components.

\subsection{Code Property Graph Construction}

To structurally represent program code as a graph, we use the \textit{Joern} static analysis framework to generate Code Property Graphs (CPGs)~\cite{cpg} for each function. A CPG integrates three foundational representations of source code: the Abstract Syntax Tree (AST), the Control Flow Graph (CFG), and the Data Flow Graph (DFG). In the resulting graph \( G = (V, E) \), each node \( v \in V \) represents a program element, such as an operator, literal, or identifier, and each directed edge \( (v_i, v_j) \in E \) encodes a semantic or syntactic relationship.

Rather than retaining the full AST, which often contains uninformative syntactic artefacts (such as punctuation or structural brackets), we include only \texttt{IS\_AST\_PARENT} edges. This decision reflects the aim of preserving structural hierarchy while avoiding noise that does not contribute meaningfully to vulnerability detection.

We define a set of 13 edge types to capture essential control, data, and structural dependencies across the code:

\begin{center}
\texttt{\{controls, declares, def, dom, flows\_to, is\_ast\_parent, is\_class\_of, is\_file\_of,} \\
\texttt{is\_function\_of\_ast, is\_function\_of\_cfg, post\_dom, reaches, use\}}
\end{center}

Each edge is categorised by its type \( t_{ij} \in \{1, 2, \dots, 13\} \), which is later embedded into a continuous vector during the learning process. Any edge that does not fall into this set is discarded. Additional graph-level preprocessing, including the removal of CPGs with invalid control flow or excessive node counts, is described in Section~\ref{subsec:dataset_and_preprocess}.

\subsection{Feature Engineering}

Each node in the Code Property Graph is represented using three complementary types of features: structural context embeddings, semantic token embeddings, and edge type encodings. Together, these modalities capture the connectivity, content, and relational dependencies of program elements, which are essential for reasoning about vulnerabilities in source code.

\subsubsection{Structural Context Embeddings}

To capture structural patterns within each Code Property Graph, we generate random walks guided by metapaths, as outlined in Algorithm~\ref{alg:metapath}. These walks traverse key relations, including control flow, data flow, and AST parent edges, in order to encode the contextual semantics of graph traversal paths. Each step in a walk is tokenised according to the following schema:

\begin{center}
\texttt{src\_type : edge\_type : tgt\_type : direction : $\Delta$depth : scope}
\end{center}

Here, \texttt{src\_type} and \texttt{tgt\_type} denote the types of the current and next nodes; \texttt{edge\_type} specifies the relation traversed; \texttt{direction} indicates forward or backward movement; $\Delta$depth represents the change in AST depth between steps; and \texttt{scope} captures higher-level contextual cues such as \texttt{return\_path} or \texttt{near\_entry}.

To learn distributed representations of these structural tokens, we train a skip-gram Word2Vec model~\cite{mikolov2013efficientestimationwordrepresentations} on the resulting sequences. Word2Vec is chosen for its ability to capture local and contextual co-occurrence patterns, allowing semantically related structures to be placed closer in the embedding space. Each node’s structural embedding is computed as the average of the embeddings of the walk tokens associated with that node. The resulting vector has a dimensionality of 512.

\subsubsection{Semantic Token Embeddings}

For semantic representation, we extract the type and content of each node from its CPG label. A node’s label typically includes a syntactic category (e.g., \texttt{CallExpression}) along with lexical tokens such as identifiers, literals, or operators. We construct a token sequence by placing the node type first, followed by content tokens extracted from the code.

To reduce vocabulary sparsity, numeric literals are replaced with a special token \texttt{NUM}. The resulting sequences are used to train a skip-gram Word2Vec model~\cite{mikolov2013efficientestimationwordrepresentations} over all training graphs. Word2Vec enables the learning of embeddings that reflect contextual similarity between tokens based on their usage patterns across code.

Each node’s semantic vector is computed by averaging the embeddings of its matched tokens. This representation allows the model to incorporate both syntactic roles and lexical content.

\subsubsection{Edge Type Encoding}

Each edge in the CPG is assigned one of 13 predefined relation types, including \texttt{controls}, \texttt{declares}, \texttt{def}, \texttt{flows\_to}, \texttt{is\_ast\_parent}, and others. These types reflect control, data, and structural dependencies crucial for code reasoning.

To represent edge types within the model, we use a trainable embedding layer that maps each edge type index to a continuous vector. The edge embedding dimension is set to 32. These embeddings are learned jointly during training, allowing the model to exploit relational differences between edges when propagating information through the graph.

\begin{algorithm}[H]
\caption{Contextual Metapath Walk Generation}
\label{alg:metapath}
\begin{algorithmic}[1]
\STATE \textsc{GenerateWalks}$(G = (V, E), L, R)$
\STATE $\mathcal{W} \gets \emptyset$
\FOR{each node $v \in V$}
    \FOR{$r = 1$ to $R$}
        \FOR{direction $\in \{\texttt{fwd}, \texttt{bwd}\}$}
            \STATE $w \gets [\,]$, \quad $c \gets v$
            \WHILE{$|w| < L$}
                \STATE $n \gets$ \textsc{RandomNeighbour}$(c, \texttt{direction})$
                \IF{$n = \emptyset$}
                    \STATE \textbf{break}
                \ENDIF
                \STATE $t \gets$ \textsc{FormatToken}$(c, n, \texttt{direction})$
                \STATE Append $t$ to $w$
                \STATE $c \gets n$
            \ENDWHILE
            \IF{$|w| \geq 2$}
                \STATE Add $w$ to $\mathcal{W}$
            \ENDIF
        \ENDFOR
    \ENDFOR
\ENDFOR
\STATE \textbf{return} $\mathcal{W}$
\end{algorithmic}
\end{algorithm}

\subsection{Edge-Aware GNN Architecture}

Let \( G = (V, E) \) denote a directed Code Property Graph, where \( V \) is the set of nodes and \( E \subseteq V \times V \) is the set of edges. Each node \( i \in V \) is associated with an initial feature vector \( \mathbf{h}_i^{(0)} \in \mathbb{R}^d \), formed by concatenating its structural and semantic embeddings. Each edge \( (i, j) \in E \) is assigned a type-specific embedding \( \mathbf{e}_{ij} \in \mathbb{R}^{d_e} \), drawn from a learned embedding table.

To enable structure-aware and relation-sensitive message passing, we employ a two-layer edge-aware GATv2 architecture~\cite{gatv2}. This depth allows the model to aggregate both first-order and second-order neighbourhood information, capturing richer context while avoiding oversmoothing and overfitting associated with deeper models.

At each layer \( l \), attention coefficients are computed between node \( i \) and each neighbour \( j \in \mathcal{N}(i) \) using the following raw attention energy:
\begin{equation}
e_{ij}^{(l)} = \mathbf{a}^\top \, \phi\left( \mathbf{W}_q \mathbf{h}_i^{(l)} \, \| \, \mathbf{W}_k \mathbf{h}_j^{(l)} \, \| \, \mathbf{W}_e \mathbf{e}_{ij} \right),
\end{equation}
where \( \phi(\cdot) \) is a non-linear activation (e.g., LeakyReLU), \( \| \) denotes vector concatenation, and \( \mathbf{W}_q, \mathbf{W}_k, \mathbf{W}_e \) are learnable projection matrices. The vector \( \mathbf{a} \) maps the concatenated representation to a scalar.

These energies are normalised using the softmax function to yield attention weights:
\begin{equation}
\alpha_{ij}^{(l)} = \frac{\exp(e_{ij}^{(l)})}{\sum_{k \in \mathcal{N}(i)} \exp(e_{ik}^{(l)})}.
\end{equation}

The updated representation of node \( i \) at layer \( l+1 \) is computed as:
\begin{equation}
\mathbf{h}_i^{(l+1)} = \sigma\left( \sum_{j \in \mathcal{N}(i)} \alpha_{ij}^{(l)} \mathbf{W}_v \mathbf{h}_j^{(l)} \right),
\end{equation}
where \( \mathbf{W}_v \) is a learnable weight matrix and \( \sigma(\cdot) \) is a non-linear activation function.

The two GATv2 layers are applied sequentially. A residual connection is added to preserve intermediate representations:
\begin{equation}
\mathbf{h}_i^{\text{out}} = \mathbf{h}_i^{(2)} + \mathbf{h}_i^{(1)}.
\end{equation}
If \( \mathbf{h}_i^{(1)} \) and \( \mathbf{h}_i^{(2)} \) differ in dimension, a linear projection is applied before summation.

To compute a graph-level embedding, we apply global attention pooling. For each node \( i \), an attention score \( \beta_i \) is computed as:
\begin{equation}
\beta_i = \frac{\exp\left( \mathbf{w}^\top \tanh\left( \mathbf{W}_g \mathbf{h}_i^{\text{out}} \right) \right)}{\sum_{j \in V} \exp\left( \mathbf{w}^\top \tanh\left( \mathbf{W}_g \mathbf{h}_j^{\text{out}} \right) \right)},
\end{equation}
where \( \mathbf{W}_g \) and \( \mathbf{w} \) are learnable parameters. The graph representation \( \mathbf{z} \) is then given by:
\begin{equation}
\mathbf{z} = \sum_{i \in V} \beta_i \mathbf{h}_i^{\text{out}}.
\end{equation}

The graph embedding is passed through a two-layer MLP~\cite{MLP} classifier. The hidden representation is:
\begin{equation}
\mathbf{h}_{\text{mlp}} = \text{ReLU}\left( \mathbf{W}_1 \mathbf{z} + \mathbf{b}_1 \right),
\end{equation}
and the final prediction is:
\begin{equation}
\hat{\mathbf{y}} = \text{softmax}\left( \mathbf{W}_2 \mathbf{h}_{\text{mlp}} + \mathbf{b}_2 \right),
\end{equation}
where \( \hat{\mathbf{y}} \in \mathbb{R}^2 \) represents the predicted class distribution. All \( \mathbf{W}_\cdot \) and \( \mathbf{b}_\cdot \) are learnable parameters.

\begin{figure*}[!t]
\centering
\includegraphics[width=5.0in]{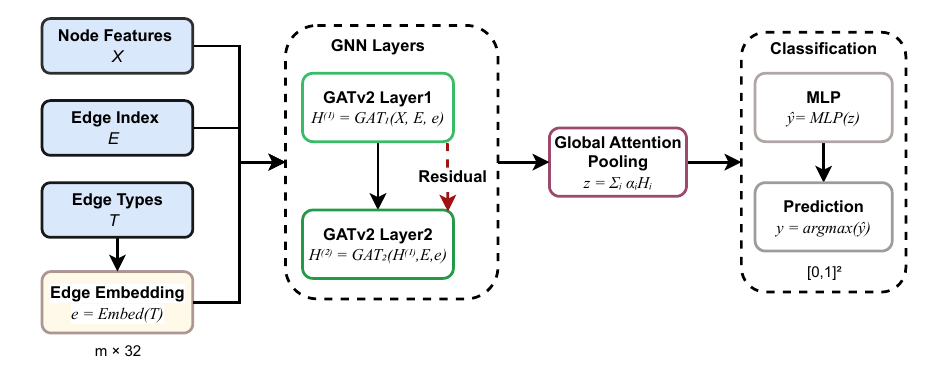}
\caption{
Overview of the ExplainVulD architecture. The model takes as input node features \( \mathbf{X} \), edge indices \( \mathbf{E} \), and edge types \( \mathbf{T} \), which are embedded and passed through two edge-aware GATv2 layers with a residual connection. Global attention pooling aggregates node-level representations into a graph-level embedding, followed by an MLP classifier. Here, \( \mathbf{H}^{(l)} = [\mathbf{h}_1^{(l)}, \dots, \mathbf{h}_n^{(l)}]^\top \) denotes the matrix of node embeddings at layer \( l \).
}
\label{model_architecture}
\end{figure*}

This architecture enables ExplainVulD to learn fine-grained representations of both local semantics and structural dependencies, improving its ability to identify subtle vulnerability patterns in real-world code graphs.

\subsection{Training and Optimisation}

Let \( \hat{\mathbf{y}} \in \mathbb{R}^2 \) denote the predicted class probability distribution for a graph, and let \( y \in \{0, 1\} \) represent the true label, where \( y = 1 \) corresponds to a vulnerable function. To account for the pronounced imbalance in the training data, we adopt a weighted cross-entropy loss that assigns greater importance to the minority class.

The loss for a single instance is defined as:
\begin{equation}
\mathcal{L}_{\text{wCE}}(\hat{\mathbf{y}}, y) = - w_y \log(p_y),
\end{equation}
where \( p_y = \hat{\mathbf{y}}[y] \) is the predicted probability of the true class, and \( w_y \in \mathbb{R}^+ \) is a class-specific weight. These weights are computed from the label distribution in the training set as follows:
\[
w_0 = 1.0, \quad w_1 = \frac{N_0 + N_1}{2N_1},
\]
where \( N_0 \) and \( N_1 \) denote the number of safe and vulnerable samples, respectively. This formulation ensures that misclassified vulnerable instances incur higher penalties, thereby mitigating the risk of minority-class underfitting.

The average loss over a batch of size \( N \) is given by:
\begin{equation}
\mathcal{L} = \frac{1}{N} \sum_{i=1}^{N} \mathcal{L}_{\text{wCE}}(\hat{\mathbf{y}}^{(i)}, y^{(i)}).
\end{equation}

Model parameters are updated using the Adam optimiser with an initial learning rate of \( 10^{-3} \) and a weight decay of \( 10^{-5} \). To improve convergence stability, the learning rate is halved whenever the validation F1 score fails to improve over two consecutive epochs:
\begin{equation}
\eta_{t+1} =
\begin{cases}
0.5 \cdot \eta_t, & \text{if validation F1 stagnates for 2 epochs}, \\
\eta_t, & \text{otherwise}.
\end{cases}
\end{equation}

To reduce overfitting, we apply dropout to the hidden layer of the classifier and monitor performance using early stopping. Training is terminated when the validation F1 score for the vulnerable class does not improve over six successive epochs.

This training procedure supports stable convergence under highly imbalanced data conditions and encourages the learning of representations that remain sensitive to the rare but critical vulnerable cases.

\subsection{Explainability of Model Predictions}

To support transparency and manual auditing in security-critical settings, we incorporate a post hoc explanation module that identifies the most influential nodes and edges in the Code Property Graph (CPG) for a given prediction. This module operates after model inference and combines internal signals,specifically attention weights and input gradients, to produce structured, fine-grained explanations of the decision process.

Let \( \mathcal{L}(\hat{y}, y) \) denote the scalar loss associated with the predicted label \( \hat{y} \) and true label \( y \). During the forward pass, we extract the final-layer node embeddings \( \mathbf{H} \in \mathbb{R}^{|V| \times d} \), edge embeddings \( \mathbf{E} \in \mathbb{R}^{|E| \times d'} \), and the global attention weights \( \boldsymbol{\alpha} \in \mathbb{R}^{|V|} \) used in pooling.

\paragraph{Node Attribution.}  
Each node \( v_i \in V \) is scored using two complementary relevance signals:

\begin{itemize}
    \item \textbf{Gradient-based relevance:} Measures the sensitivity of the model’s output to perturbations in the node's input features. For class \( c \), we compute:
    \begin{equation}
        r_i^{\text{(grad)}} = \left\| \nabla_{\mathbf{x}_i} \hat{y}_c \right\|_2.
    \end{equation}

    \item \textbf{Attention-based relevance:} The global attention weight \( \alpha_i \), assigned during graph-level pooling, reflects the node’s contribution to the overall output.
\end{itemize}

The final node relevance score is the mean of the two components:
\begin{equation}
    r_i = \frac{1}{2} \left( r_i^{\text{(grad)}} + \alpha_i \right).
\end{equation}

\paragraph{Edge Attribution.}  
Each edge \( e_{ij} \in E \) is similarly scored using:

\begin{itemize}
    \item \textbf{Gradient-based relevance:} The sensitivity of the output to the edge’s learned embedding:
    \begin{equation}
        s_{ij}^{\text{(grad)}} = \left\| \nabla_{\mathbf{e}_{ij}} \hat{y}_c \right\|_2.
    \end{equation}

    \item \textbf{Embedding norm proxy:} The \( \ell_2 \)-norm of the edge embedding provides a heuristic for its structural importance:
    \begin{equation}
        s_{ij}^{\text{(emb)}} = \left\| \mathbf{e}_{ij} \right\|_2.
    \end{equation}
\end{itemize}

The final edge relevance score is given by:
\begin{equation}
    s_{ij} = \frac{1}{2} \left( s_{ij}^{\text{(grad)}} + s_{ij}^{\text{(emb)}} \right).
\end{equation}

\paragraph{Explanation Output and Interpretation.}  
After computing \( \{r_i\} \) and \( \{s_{ij}\} \), the module ranks all nodes and edges and extracts the top-\( k \) most influential elements. These are mapped back to their original CPG components, including node labels and edge types, and visualised as a highlighted subgraph.

This procedure provides both a quantitative explanation of the model’s internal decision and a human-readable interpretation in terms of relevant code structures and dependencies. Such explanations can assist developers and security analysts in understanding model behaviour, verifying results, and supporting downstream auditing.

\section{Experimental Setup}
\label{sec:experimental_setup}

This section describes the dataset and preprocessing, evaluation metrics, evaluation procedure, baseline methods, research questions, and implementation details used to assess the performance of the proposed framework.

\subsection{Dataset and Preprocessing}
\label{subsec:dataset_and_preprocess}

We evaluate the proposed ExplainVulD framework using the ReVeal~\cite{chakraborty2021deep} dataset, which contains vulnerable and non-vulnerable C/C++ functions extracted from historical vulnerability-fix commits in two large, actively maintained open-source projects: Chromium and the Linux Debian Kernel. These projects were chosen due to their long development histories, wide deployment, and the availability of publicly documented vulnerability reports.

ReVeal was constructed by identifying commits tagged with security labels in each project's respective vulnerability tracker. For each commit, both the pre-patch (vulnerable) and post-patch (fixed) versions of modified functions were extracted. Additionally, any functions within the same file that were not modified by the patch were included as safe. This approach results in function-level annotations that more closely resemble how vulnerability prediction would be deployed in practice, where the goal is to flag specific functions within large codebases~\cite{chakraborty2021deep}.

The dataset reflects real-world imbalance: only a small proportion of functions are vulnerable, while the majority are safe. This makes the detection task more challenging and realistic, as models must identify subtle patterns associated with vulnerabilities without being overwhelmed by the dominant safe class. The diversity in code structure, naming, and style across the two source projects further increases the complexity of generalisation.

Compared to prior benchmarks, ReVeal offers a more realistic evaluation setting for function-level vulnerability detection. SARD~\cite{black2017sard} is fully synthetic and was developed primarily to benchmark static analysis tools. It lacks the structural diversity and complexity of production code. Draper~\cite{russell2018automated} consists of semi-synthetic samples created by mutating and compiling code fragments, but includes limited real-world code. The FFmpeg+QEMU subset used in Devign~\cite{zhou2019devign} consists of functions from real projects, but assumes an approximately balanced class distribution, with roughly 45\% labelled as vulnerable. This does not reflect the sparsity of vulnerabilities in real systems. Moreover, it includes only those functions directly modified by security patches and excludes other functions from the same context, restricting the model’s ability to differentiate vulnerable functions from nearby safe ones.

ReVeal addresses these limitations by including all functions modified in security patches as well as their fixed counterparts and surrounding unmodified code from the same file~\cite{chakraborty2021deep}. This design preserves the local context in which vulnerabilities occur and allows models to learn how vulnerable functions differ not only from their repaired forms but also from adjacent safe code. Such contextual contrast is critical for robust detection in practical scenarios.

As part of preprocessing, we exclude any Code Property Graph (CPG) with more than 500 nodes to reduce noise and prevent training instability. We also remove functions that do not contain a valid Control Flow Graph (CFG), as control-flow information is essential for capturing execution behaviour relevant to vulnerability detection.

\renewcommand{\arraystretch}{1.15}
\begin{table}[ht]
\centering
\caption{Distribution of functions in the ReVeal dataset before data cleaning}
\label{tab:reveal_before_cleaning}
\begin{tabular}{l|r|r|r}
\hline
\textbf{Project} & \textbf{Vulnerable} & \textbf{Safe} & \textbf{Vulnerable (\%)} \\
\hline
Chrome & 825 & 3,611 & 22.85 \\
Debian & 1,415 & 16,883 & 8.38 \\
\hline
\textbf{Total} & 2,240 & 20,494 & 10.93 \\
\hline
\end{tabular}
\end{table}
\renewcommand{\arraystretch}{1.15}
\begin{table}[ht]
\centering
\caption{Distribution of functions in the ReVeal dataset after data cleaning}
\label{tab:reveal_after_cleaning}
\begin{tabular}{l|r|r|r}
\hline
\textbf{Project} & \textbf{Vulnerable} & \textbf{Safe} & \textbf{Vulnerable (\%)} \\
\hline
Chrome & 577 & 2,377 & 19.53 \\
Debian & 896 & 13,861 & 6.06 \\
\hline
\textbf{Total} & 1,473 & 16,238 & 8.32 \\
\hline
\end{tabular}
\end{table}

\noindent\textbf{Note:} The cleaned dataset containing code functions and CPGs are publicly available at \url{https://github.com/Radowan98/ExplainVulD} to support reproducibility and future research.

\subsection{Evaluation Metrics}

To evaluate the performance of our framework, we use a set of standard classification metrics commonly applied in imbalanced learning scenarios. These include Accuracy, Precision, Recall, F1-score, and the Area Under the Receiver Operating Characteristic Curve (AUC-ROC). While Accuracy offers a high-level view of overall correctness, it can be misleading in imbalanced settings. Therefore, we place greater emphasis on Precision, Recall, F1, and AUC, which better capture the model’s effectiveness in identifying vulnerable functions.

Given the security-critical nature of the task, we report metrics with a focus on the positive class, allowing for a more meaningful assessment of the model's ability to detect true vulnerabilities under real-world conditions.

\subsection{Evaluation Procedure}
Deep learning models are inherently sensitive to various sources of randomness~\cite{10.5555/249049}, including weight initialization, dropout, and data ordering. Furthermore, our dataset is randomly split into 80\% training, 10\% validation, and 10\% test sets for each run. To ensure robust and unbiased evaluation, we repeat the entire training and evaluation process 30 times with different random splits. Final results are reported as the mean and standard deviation across these runs. This prevents cherry-picking and aligns with best practices adopted by prior work such as ReVeal~\cite{chakraborty2021deep}. 

As Flawfinder~\cite{wheeler2025flawfinder}, and Cppcheck~\cite{cppcheck} are static analysis tools and do not involve random initialisation or training, their output remains fixed for a given input. However, its performance can still vary across different types of functions and vulnerability classes. To enable a fair comparison, we evaluate Flawfinder and Cppcheck on the same 30 randomly sampled test splits used for our model. This allows us to assess the consistency and limitations of static analysis with data variability.

\subsection{Baselines}

We compare our proposed framework, ExplainVulD, against ReVeal~\cite{chakraborty2021deep}, a graph-based vulnerability detection method that achieves significantly higher F1-scores than earlier approaches such as Devign~\cite{zhou2019devign}, SySeVR~\cite{9321538-sysevr}, VulDeePecker~\cite{zou2019uvuldeepecker}, and the work by Russell et al.~\cite{russell2018automated}. As this advantage is well established in prior work, we adopt ReVeal as our primary learning-based baseline.

In addition, we benchmark against two static analysis tools: Flawfinder~\cite{wheeler_flawfinder} and Cppcheck~\cite{cppcheck}. These tools are suitable for our dataset, which consists of standalone C functions without header files or full project context. This constraint excludes static analysers such as Infer~\cite{infer2025}, which depend on complete build environments or compilation units.

Flawfinder performs rule-based analysis by matching code against a database of known insecure functions and patterns~\cite{wheeler_flawfinder}. Cppcheck performs flow-sensitive analysis by tracking control and data flow to detect issues such as buffer overflows, memory misuse, and uninitialised variables~\cite{cppcheck}. They support both C and C++ and produce structured diagnostic reports.

These two tools represent complementary approaches to static analysis and serve as practical, compiler-independent baselines. Their inclusion enables a comprehensive evaluation of ExplainVulD relative to established techniques used in real-world software engineering.

\subsection{Research Questions}
\label{rqs}
To evaluate the proposed framework, we define the following research questions:
\begin{itemize}
    \item \textbf{RQ1:} How accurately and reliably does ExplainVulD detect software vulnerabilities in real-world, imbalanced C/C++ code?

    \item \textbf{RQ2:} How do different imbalance-handling strategies, including focal loss and class weighting, affect detection performance and training stability?

    \item \textbf{RQ3:} What is the contribution of each feature component (semantic embedding, structural embedding, edge-type encoding) to overall model performance?

   \item \textbf{RQ4:} How does ExplainVulD compare to existing vulnerability detection methods, including prior learning-based models and static analysis tools?

    \item \textbf{RQ5:} How explainable are ExplainVulD’s predictions, and do the generated explanations accurately identify the code regions that influence the model’s classification decisions?

\end{itemize}

\subsection{Implementation Details}
All experiments are conducted on a Google Cloud Platform (GCP) virtual machine equipped with an NVIDIA Tesla T4 GPU (16GB VRAM), 8 vCPUs, and 52 GB RAM, running Ubuntu 20.04. The model is trained using the Adam optimiser with a learning rate of 0.001 and a batch size of 32. Class-weighted cross-entropy loss is employed to address the severe class imbalance. Word2Vec models are trained using the Gensim library. The entire framework is implemented in Python 3.10 using PyTorch 2.1 and PyTorch Geometric (PyG) 2.4. The edge-aware GATv2 architecture consists of two attention layers.

\begin{table}[h!]
\centering
\caption{Implementation Details of Embedding and Model Components}
\label{tab:implementation-details}
\begin{tabular}{|l|l|}
\hline
\textbf{Component} & \textbf{Configuration} \\
\hline
\multicolumn{2}{|c|}{\textit{Semantic Word2Vec}} \\
\hline
Vector size & 512 \\
Window size & 5 \\
Epochs & 80 \\
Minimum count & 1 \\
Skip-gram (SG) & 1 \\
Negative sampling & 15 \\
\hline
\multicolumn{2}{|c|}{\textit{Structural Word2Vec }} \\
\hline
Walk length & 20 \\
Number of walks per node & 10 \\
Vector size & 512 \\
Window size & 5 \\
Epochs & 120 \\
Minimum count & 1 \\
Skip-gram (SG) & 1 \\
Negative sampling & 15 \\
\hline
\multicolumn{2}{|c|}{\textit{Edge-Aware GATv2 Model}} \\
\hline
Input dimension & 1024 (concatenated semantic + structural) \\
Edge embedding dimension & 32 \\
Number of layers & 2 \\
Epochs & 100 \\
Early stopping patience & 6 \\
\hline
\end{tabular}
\end{table}


\section{Results and Analysis}
\label{sec:results}

We present our findings in alignment with the research questions introduced in Section ~\ref{rqs}. Each subsection below addresses one research question through relevant experiments, metrics, and observations.
In this section, we will answer the research questions.

\subsection{Accuracy and Reliability of the Proposed Framework (RQ1)}

To assess the accuracy and reliability of ExplainVulD, we conducted 30 independent runs with randomly sampled training, validation, and test splits in an 80/10/10 ratio. This evaluation protocol reduces sensitivity to initialisation and data partitioning, supporting reproducibility. Using class-weighted cross-entropy loss, the model achieved an average accuracy of 88.25\%, F1 score of 48.23\%, and AUC of 87.64\%. Mean precision and recall were 38.16\% and 66.09\%, respectively. Standard deviations were low across metrics, including 0.84\% for accuracy, 3.09\% for F1 score, and 2.00\% for AUC. The best-performing run recorded an F1 score of 51.37\%, recall of 73.96\%, and AUC of 90.68\%. The maximum accuracy and precision observed were 89.83\% and 42.26\%, respectively. These results suggest the model can be configured to prioritise different objectives, such as recall, depending on deployment needs.

\begin{figure*}[!t]
\centering

\subfloat[Accuracy]{\includegraphics[width=0.3\textwidth]{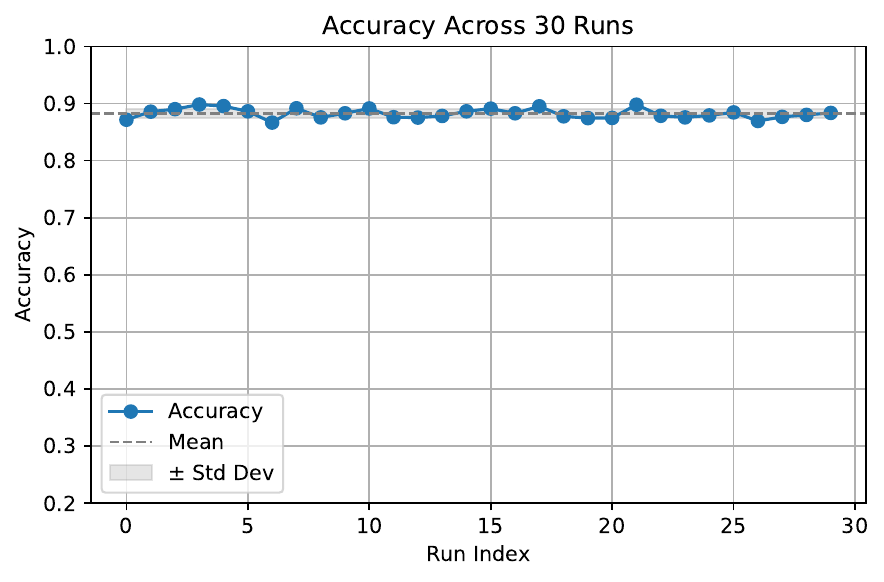}\label{fig:rq2-acc}}
\hspace{6pt}
\subfloat[Precision]{\includegraphics[width=0.3\textwidth]{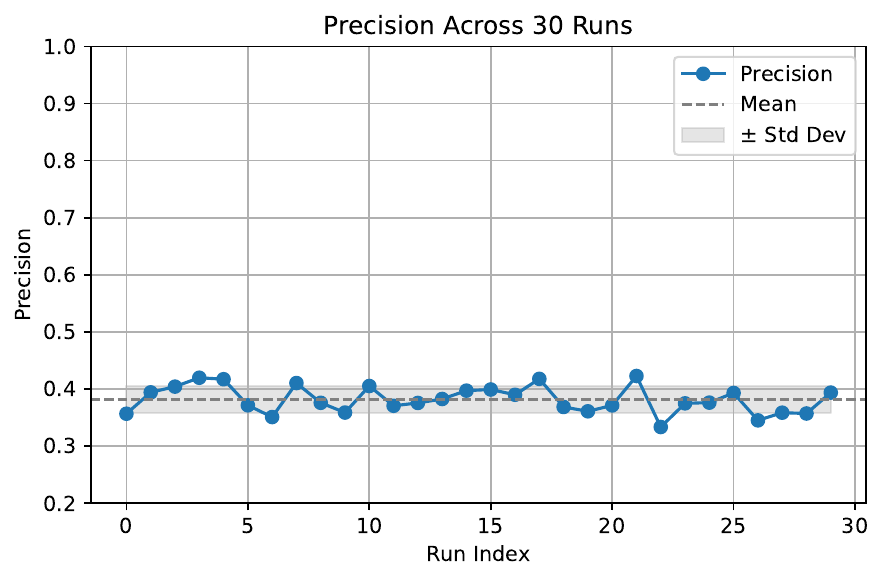}\label{fig:rq2-prec}}
\hspace{6pt}
\subfloat[Recall]{\includegraphics[width=0.3\textwidth]{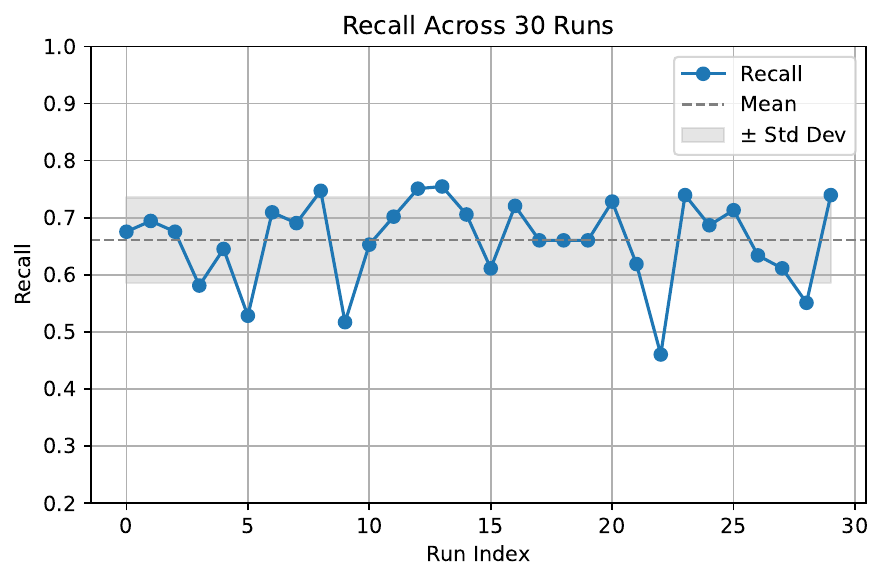}\label{fig:rq2-recall}}

\vspace{6pt}

\subfloat[F1 Score]{\includegraphics[width=0.3\textwidth]{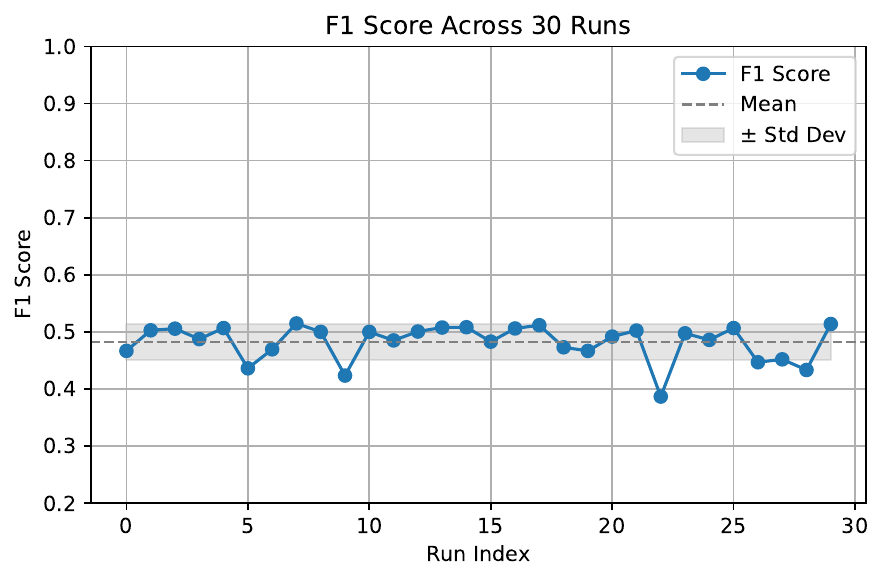}\label{fig:rq2-f1}}
\hspace{6pt}
\subfloat[AUC]{\includegraphics[width=0.3\textwidth]{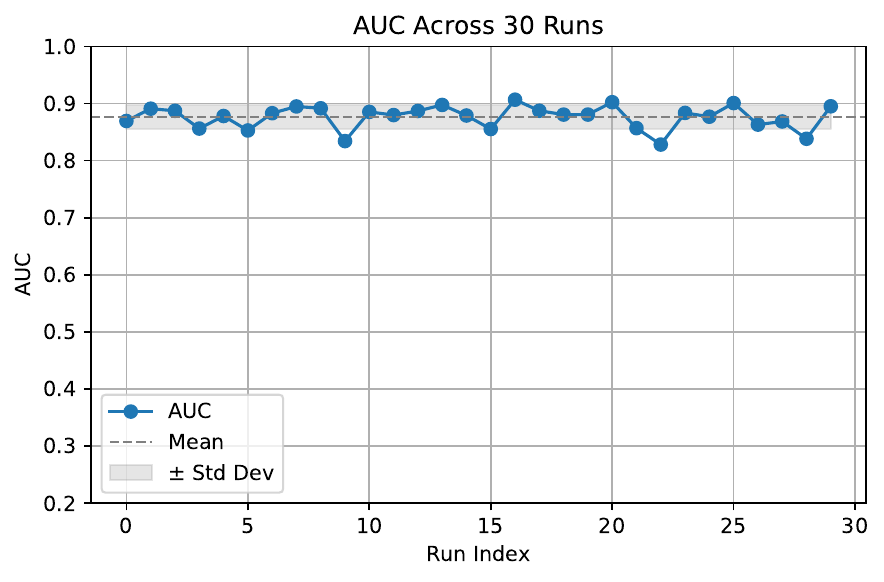}\label{fig:rq2-auc}}

\caption{Performance across 30 runs using class-weighted cross-entropy. Each subplot shows per-run values, the mean (dashed line), and $\pm$ standard deviation (shaded area), illustrating consistency across metrics.}
\label{fig:rq1-focal}
\end{figure*}

Table~\ref{tab:focal-summary} reports the mean, variance, and best values across all runs. Figure~\ref{fig:rq1-focal} shows the per-run metric distributions, with shaded bands indicating one standard deviation around the mean. Overall, the results indicate that the proposed framework maintains consistent performance across multiple runs on a real-world imbalanced dataset. The use of class-weighted cross-entropy supports stable learning across classes without requiring fine-grained tuning.

\begin{table}[!t]
\centering
\caption{Mean, standard deviation, and best score across 30 runs using class-weighted cross-entropy.}
\label{tab:focal-summary}
\begin{tabular}{|l|c|c|c|}
\hline
\textbf{Metric} & \textbf{Mean (\%)} & \textbf{Std Dev (\%)} & \textbf{Best (\%)} \\
\hline
Accuracy  & 88.25 & 0.84 & 89.83 \\
Precision & 38.16 & 2.35 & 42.26 \\
Recall    & 66.09 & 7.33 & 73.96 \\
F1 Score  & 48.23 & 3.09 & 51.37 \\
AUC       & 87.64 & 2.00 & 90.68 \\
\hline
\end{tabular}
\end{table}

\subsection{Class Imbalance Strategies (RQ2)}

This section evaluates how different loss functions affect performance under severe class imbalance, a common characteristic in vulnerability detection datasets. We compare five focal loss configurations against a class-weighted cross-entropy baseline, evaluating their impact on accuracy, recall, F1 score, and AUC over 30 independent runs.

Focal loss introduces two tunable parameters: $\alpha$, which adjusts the relative importance of positive and negative examples, and $\gamma$, which down-weights easy examples to focus learning on hard or misclassified cases. Table~\ref{tab:focal-loss-params} summarises the rationale behind each configuration.

\begin{table}[h!]
\centering
\caption{Combinations of focal loss parameters $\alpha$ and $\gamma$ for highly imbalanced vulnerability detection.}
\label{tab:focal-loss-params}
\begin{tabular}{|c|c|p{6.8cm}|}
\hline
\textbf{$\alpha$} & \textbf{$\gamma$} & \textbf{Reasoning} \\
\hline
1.0  & 2.0 & Standard setting proposed by Lin et al.~\cite{focal}. \\
\hline
0.95 & 2.5 & Balances hard-to-classify and minority-class examples. \\
\hline
0.90 & 2.5 & Increases attention on difficult samples while retaining balance. \\
\hline
0.85 & 3.0 & Suppresses well-classified cases, promoting minority-class learning. \\
\hline
0.80 & 3.0 & Maximises focus on hard and underrepresented instances. \\
\hline
\end{tabular}
\end{table}

\begin{table}[!t]
\centering
\caption{Mean and standard deviation (in parentheses, \% scale) of performance metrics across 30 runs for each focal loss configuration and class-weighted cross-entropy.}
\label{tab:focal-metrics}
\begin{tabular}{|p{2.7cm}|c|c|c|c|c|}
\hline
\textbf{Configuration} 
  & \textbf{Accuracy} 
  & \textbf{Precision} 
  & \textbf{Recall} 
  & \textbf{F1 Score} 
  & \textbf{AUC} \\
\hline
$(\alpha{=}0.80,\ \gamma{=}3.0)$ 
  & \begin{tabular}{@{}c@{}}58.3 \\ \scriptsize(3.1)\end{tabular} 
  & \begin{tabular}{@{}c@{}}\textbf{53.5} \\ \scriptsize(2.7)\end{tabular} 
  & \begin{tabular}{@{}c@{}}\textbf{72.8} \\ \scriptsize(9.3)\end{tabular} 
  & \begin{tabular}{@{}c@{}}\textbf{61.2} \\ \scriptsize(2.0)\end{tabular} 
  & \begin{tabular}{@{}c@{}}65.1 \\ \scriptsize(2.4)\end{tabular} \\
\hline
$(\alpha{=}0.85,\ \gamma{=}3.0)$ 
  & \begin{tabular}{@{}c@{}}\textbf{91.9} \\ \scriptsize(0.7)\end{tabular} 
  & \begin{tabular}{@{}c@{}}52.0 \\ \scriptsize(5.2)\end{tabular} 
  & \begin{tabular}{@{}c@{}}42.7 \\ \scriptsize(5.0)\end{tabular} 
  & \begin{tabular}{@{}c@{}}46.6 \\ \scriptsize(3.4)\end{tabular} 
  & \begin{tabular}{@{}c@{}}86.8 \\ \scriptsize(1.3)\end{tabular} \\
\hline
$(\alpha{=}0.90,\ \gamma{=}2.5)$ 
  & \begin{tabular}{@{}c@{}}91.5 \\ \scriptsize(0.9)\end{tabular} 
  & \begin{tabular}{@{}c@{}}49.7 \\ \scriptsize(6.0)\end{tabular} 
  & \begin{tabular}{@{}c@{}}42.6 \\ \scriptsize(6.2)\end{tabular} 
  & \begin{tabular}{@{}c@{}}45.4 \\ \scriptsize(4.0)\end{tabular} 
  & \begin{tabular}{@{}c@{}}87.0 \\ \scriptsize(2.2)\end{tabular} \\
\hline
$(\alpha{=}0.95,\ \gamma{=}2.5)$ 
  & \begin{tabular}{@{}c@{}}91.6 \\ \scriptsize(0.7)\end{tabular} 
  & \begin{tabular}{@{}c@{}}49.8 \\ \scriptsize(5.5)\end{tabular} 
  & \begin{tabular}{@{}c@{}}42.0 \\ \scriptsize(6.9)\end{tabular} 
  & \begin{tabular}{@{}c@{}}45.1 \\ \scriptsize(4.5)\end{tabular} 
  & \begin{tabular}{@{}c@{}}86.6 \\ \scriptsize(2.1)\end{tabular} \\
\hline
$(\alpha{=}1.00,\ \gamma{=}2.0)$ 
  & \begin{tabular}{@{}c@{}}91.8 \\ \scriptsize(0.8)\end{tabular} 
  & \begin{tabular}{@{}c@{}}52.6 \\ \scriptsize(6.1)\end{tabular} 
  & \begin{tabular}{@{}c@{}}40.1 \\ \scriptsize(8.4)\end{tabular} 
  & \begin{tabular}{@{}c@{}}44.6 \\ \scriptsize(5.0)\end{tabular} 
  & \begin{tabular}{@{}c@{}}86.4 \\ \scriptsize(2.1)\end{tabular} \\
\hline
Class-Weighted CE
  & \begin{tabular}{@{}c@{}}88.2 \\ \scriptsize(0.8)\end{tabular} 
  & \begin{tabular}{@{}c@{}}38.2 \\ \scriptsize(2.4)\end{tabular} 
  & \begin{tabular}{@{}c@{}}66.1 \\ \scriptsize(7.3)\end{tabular} 
  & \begin{tabular}{@{}c@{}}48.2 \\ \scriptsize(3.1)\end{tabular} 
  & \begin{tabular}{@{}c@{}}\textbf{87.7} \\ \scriptsize(2.0)\end{tabular} \\
\hline
\end{tabular}
\end{table}

The configuration $(\alpha = 0.80, \gamma = 3.0)$ achieved the highest recall (72.8\%) and F1 score (61.2\%), but suffered from low accuracy (58.3\%) and AUC (65.1\%). This indicates that while the model was highly sensitive to vulnerable instances, it misclassified many safe samples, resulting in weaker generalisation. By contrast, $(\alpha = 0.85, \gamma = 3.0)$ produced the highest accuracy (91.9\%) and strong overall balance, but did not outperform the class-weighted baseline on F1 or AUC. Other configurations yielded moderate improvements but required careful parameter tuning and did not offer a consistent advantage.

Overall, although focal loss can boost specific metrics when tuned carefully, it is sensitive to configuration and may introduce instability. The class-weighted cross-entropy baseline achieved the most balanced results across all criteria: an F1 score of 48.2\%, AUC of 87.7\%, and recall of 66.1\%. It also maintained low variance across runs.

Importantly, class-weighted cross-entropy does not require additional hyperparameter tuning, making it a more practical and robust choice for deployment. It promotes consistent learning across both classes without amplifying noise from minority examples. Based on these findings, we adopt the class-weighted loss function in the final configuration of the model.

\begin{figure*}[!t]
\centering

\subfloat[Accuracy]{\includegraphics[width=0.3\textwidth]{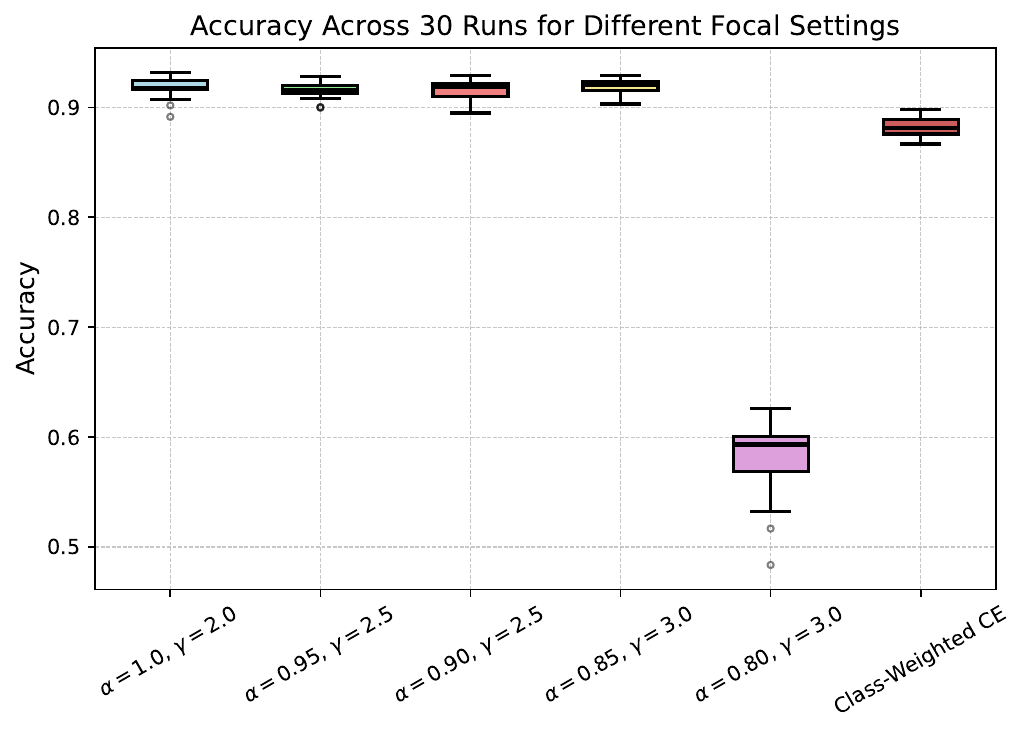}\label{fig:box-acc}}
\hspace{6pt}
\subfloat[Precision]{\includegraphics[width=0.3\textwidth]{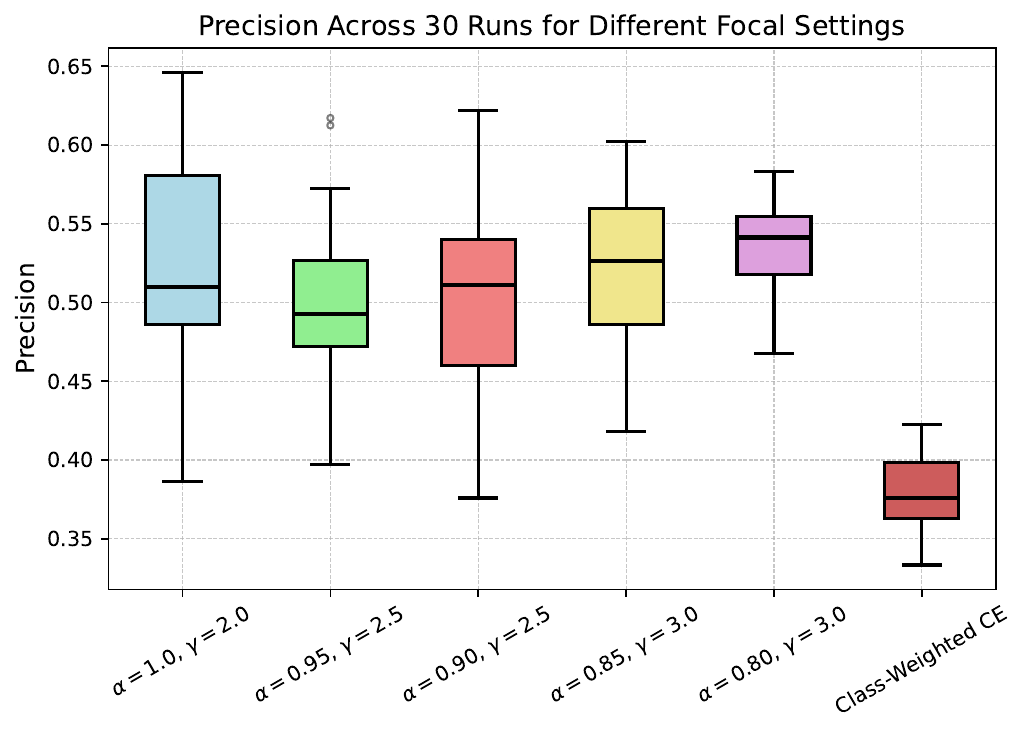}\label{fig:box-prec}}
\hspace{6pt}
\subfloat[Recall]{\includegraphics[width=0.3\textwidth]{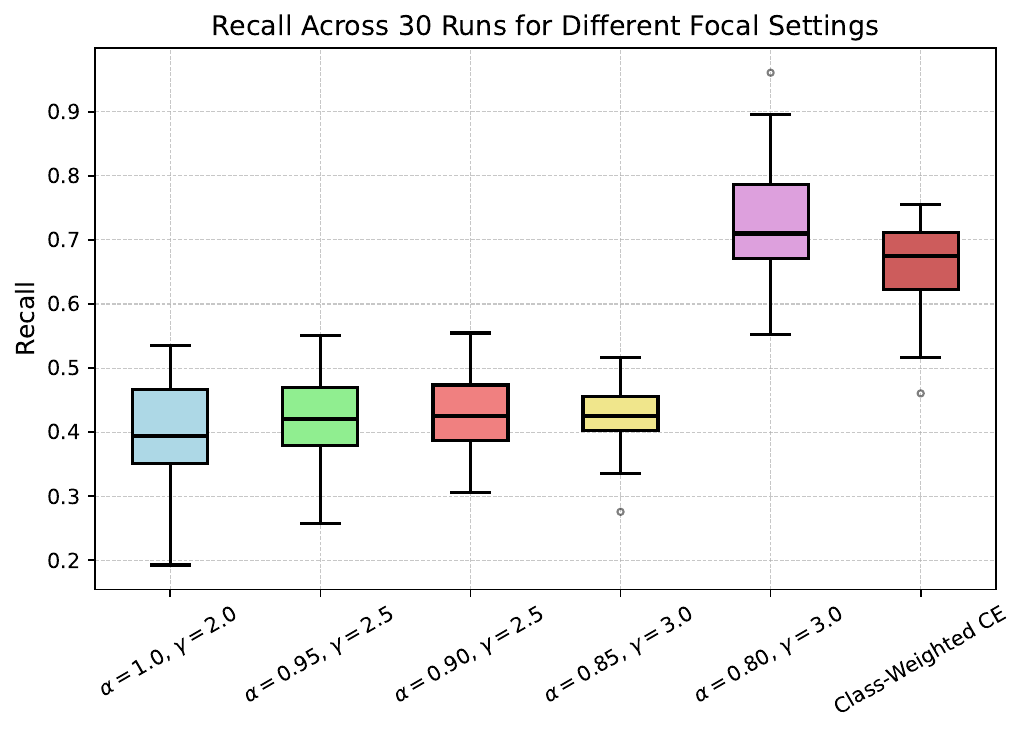}\label{fig:box-recall}}

\vspace{6pt}

\subfloat[F1 Score]{\includegraphics[width=0.3\textwidth]{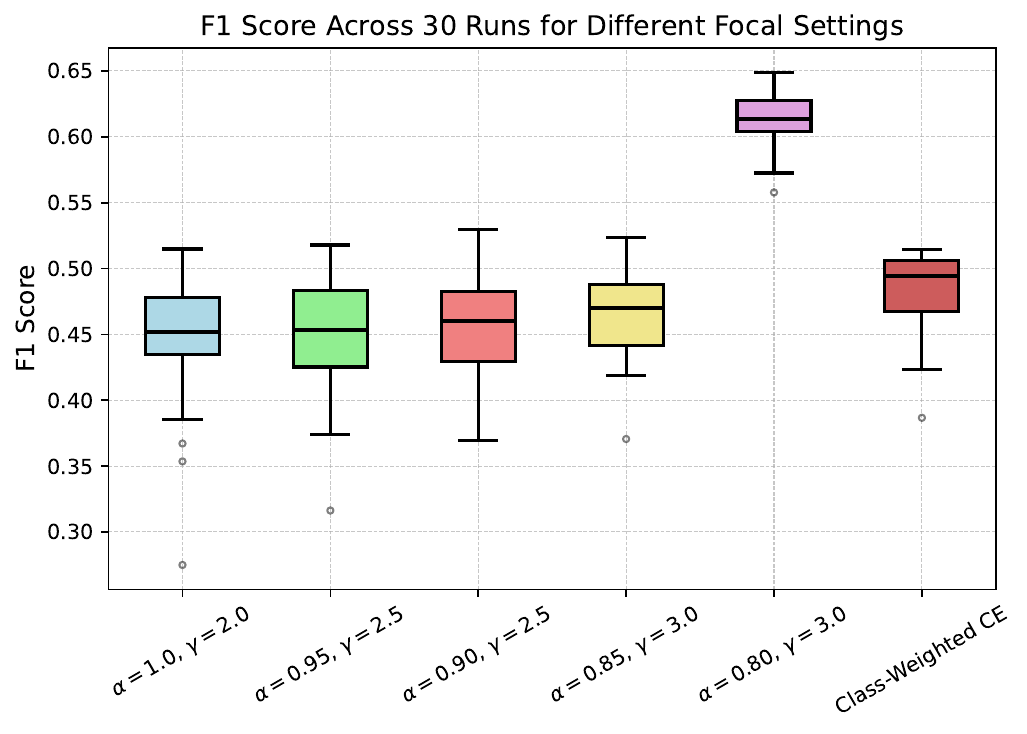}\label{fig:box-f1}}
\hspace{6pt}
\subfloat[AUC]{\includegraphics[width=0.3\textwidth]{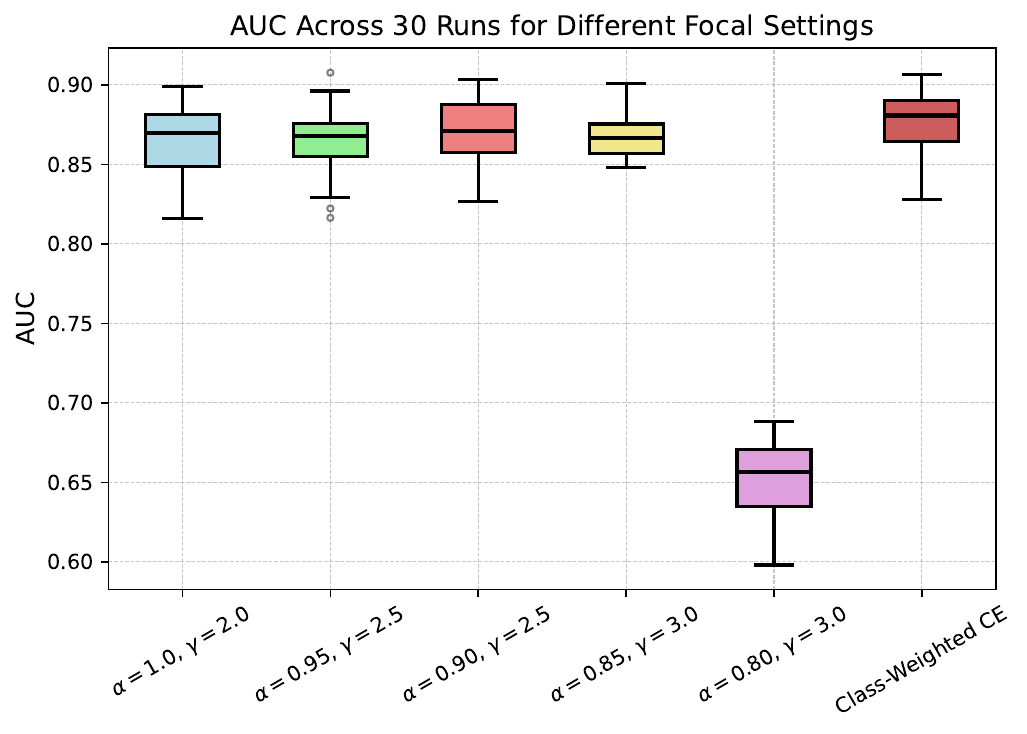}\label{fig:box-auc}}

\caption{Performance metrics across 30 runs for different focal loss settings and class-weighted CE loss. Each subplot compares configurations using varying $\alpha$ and $\gamma$ values and class-weighted CE loss, showing distribution, median, and variability.}
\label{fig:boxplots-focal}
\end{figure*}

\subsection{Feature Contribution (RQ3)}
\label{sec:feature-contribution}

Table~\ref{tab:rq3_results} presents an ablation study assessing the individual and combined contributions of semantic, structural, and edge-type features to model performance.

Using only semantic embeddings, which encode token and type information from filtered AST nodes, results in high accuracy (80.15\%) but substantially lower precision (19.91\%) and F1 score (27.01\%). This outcome reflects the model’s tendency to favour the majority class when relying solely on local syntactic information, with limited capacity to identify vulnerable functions in an imbalanced setting.

In contrast, structural embeddings trained on metapath-guided random walks yield consistently better performance. When used without edge-type information, they achieve an F1 score of 41.37\% and an AUC of 83.11\%. These results suggest that structural features provide valuable context by capturing control, data, and dominance patterns across the graph.

Adding edge-type embeddings to structural features offers a modest improvement, increasing the F1 score to 41.69\% and the AUC to 83.43\%. While this effect is limited in isolation, it becomes more significant when combined with semantic information.

Combining semantic and structural embeddings without edge-type encodings leads to further gains, with an F1 score of 44.65\% and an AUC of 86.21\%. This demonstrates that the two node-level modalities are complementary: semantic features anchor the representation in lexical content, while structural features provide global code context.

The best results are observed when all three feature types are used together. The configuration incorporating semantic and structural embeddings alongside edge-type encodings achieves an F1 score of 48.23\% and an AUC of 87.64\%. These gains indicate that edge-type information enhances message passing by distinguishing between different program relations such as AST, CFG, and DFG edges.

In summary, semantic embeddings contribute local content, structural embeddings capture contextual flow, and edge-type features enable relational reasoning. The combination of all three modalities leads to a more expressive graph representation and supports more effective vulnerability detection.

\begin{table}[!t]
\centering
\caption{Mean and standard deviation (in parentheses, \% scale) of evaluation metrics by feature modality, with and without edge-type embeddings.}
\label{tab:rq3_results}
\small
\begin{tabular}{|p{1.6cm}|r|r|r|r|r|}
\hline
\textbf{Modality} 
  & \textbf{Acc.} 
  & \textbf{Prec.} 
  & \textbf{Rec.} 
  & \textbf{F1} 
  & \textbf{AUC} \\
\hline 
sem.+strct.* 
  & \begin{tabular}{@{}c@{}}\textbf{88.25}\\\scriptsize(0.84)\end{tabular} 
  & \begin{tabular}{@{}c@{}}\textbf{38.16}\\\scriptsize(2.35)\end{tabular} 
  & \begin{tabular}{@{}c@{}}\textbf{66.09}\\\scriptsize(7.33)\end{tabular} 
  & \begin{tabular}{@{}c@{}}\textbf{48.23}\\\scriptsize(3.09)\end{tabular} 
  & \begin{tabular}{@{}c@{}}\textbf{87.64}\\\scriptsize(2.00)\end{tabular} \\
\hline
sem.* 
  & \begin{tabular}{@{}c@{}}80.15\\\scriptsize(3.49)\end{tabular} 
  & \begin{tabular}{@{}c@{}}19.91\\\scriptsize(3.23)\end{tabular} 
  & \begin{tabular}{@{}c@{}}44.12\\\scriptsize(8.31)\end{tabular} 
  & \begin{tabular}{@{}c@{}}27.01\\\scriptsize(3.12)\end{tabular} 
  & \begin{tabular}{@{}c@{}}71.39\\\scriptsize(2.28)\end{tabular} \\
\hline
struct.* 
  & \begin{tabular}{@{}c@{}}86.72\\\scriptsize(1.55)\end{tabular} 
  & \begin{tabular}{@{}c@{}}32.96\\\scriptsize(3.16)\end{tabular} 
  & \begin{tabular}{@{}c@{}}56.35\\\scriptsize(5.81)\end{tabular} 
  & \begin{tabular}{@{}c@{}}41.37\\\scriptsize(2.88)\end{tabular} 
  & \begin{tabular}{@{}c@{}}83.11\\\scriptsize(1.67)\end{tabular} \\
\hline
sem.+strct.† 
  & \begin{tabular}{@{}c@{}}87.41\\\scriptsize(1.66)\end{tabular} 
  & \begin{tabular}{@{}c@{}}35.50\\\scriptsize(4.34)\end{tabular} 
  & \begin{tabular}{@{}c@{}}61.22\\\scriptsize(8.55)\end{tabular} 
  & \begin{tabular}{@{}c@{}}44.65\\\scriptsize(4.69)\end{tabular} 
  & \begin{tabular}{@{}c@{}}86.21\\\scriptsize(2.51)\end{tabular} \\
\hline
sem.† 
  & \begin{tabular}{@{}c@{}}79.96\\\scriptsize(3.19)\end{tabular} 
  & \begin{tabular}{@{}c@{}}19.54\\\scriptsize(2.66)\end{tabular} 
  & \begin{tabular}{@{}c@{}}44.24\\\scriptsize(8.79)\end{tabular} 
  & \begin{tabular}{@{}c@{}}26.74\\\scriptsize(2.79)\end{tabular} 
  & \begin{tabular}{@{}c@{}}70.98\\\scriptsize(2.25)\end{tabular} \\
\hline
struct.† 
  & \begin{tabular}{@{}c@{}}86.72\\\scriptsize(1.86)\end{tabular} 
  & \begin{tabular}{@{}c@{}}33.32\\\scriptsize(4.30)\end{tabular} 
  & \begin{tabular}{@{}c@{}}56.78\\\scriptsize(5.01)\end{tabular} 
  & \begin{tabular}{@{}c@{}}41.69\\\scriptsize(2.94)\end{tabular} 
  & \begin{tabular}{@{}c@{}}83.43\\\scriptsize(1.96)\end{tabular} \\
\hline
\end{tabular}
\vspace{1mm}
\begin{flushleft}
\footnotesize
\textbf{*} Includes edge-type embeddings. \\
\textbf{†} Excludes edge-type embeddings to isolate node-level feature effects.
\end{flushleft}
\end{table}

\subsection{Comparative Effectiveness (RQ4)}

\begin{table}[!t]
\caption{Mean and standard deviation (in parentheses, \% scale) of performance metrics across 30 runs for each method.}
\label{tab:comparison}
\centering
\renewcommand{\arraystretch}{1.2}
\begin{tabular}{|l|r|r|r|r|}
\hline
\textbf{Method} 
& \textbf{Accuracy} 
& \textbf{Precision} 
& \textbf{Recall} 
& \textbf{F1 Score} \\
\hline
Cppcheck~\cite{cppcheck}
  & 77.34 (8.00) 
  & 18.76 (3.87) 
  & 24.04 (7.28) 
  & 20.76 (4.60) \\
Flawfinder~\cite{wheeler_flawfinder}
  & 79.56 (2.27) 
  & 18.71 (7.01) 
  & 18.44 (8.85) 
  & 16.01 (3.65) \\
ReVeal~\cite{chakraborty2021deep}
  & 84.37 (1.73) 
  & 30.91 (2.76) 
  & 60.91 (7.89)
  & 41.25 (2.28) \\
ExplainVulD (Proposed) 
  & \textbf{88.25 (0.84)} 
  & \textbf{38.16 (2.35)} 
  & \textbf{66.09 (7.33)} 
  & \textbf{48.23 (2.00)} \\
\hline
\end{tabular}
\end{table}

Table~\ref{tab:comparison} presents a comparative analysis of the proposed framework, two static analysis tools (Cppcheck and Flawfinder), and the learning-based ReVeal model~\cite{chakraborty2021deep}. The proposed method achieves the highest scores across all reported metrics, demonstrating superior performance in vulnerability detection tasks.

Compared to the static tools, the gap is substantial. Cppcheck and Flawfinder report F1 scores of 20.76\% and 16.01\%, respectively, due to their low precision and recall. These results reflect the limitations of pattern-matching heuristics when applied to real-world, imbalanced datasets without contextual reasoning. Although these tools operate without the need for project-specific configurations or learning, their effectiveness remains limited for fine-grained detection at the function level.

Against ReVeal, a strong learning-based baseline, the framework achieves a 3.9 percentage point improvement in accuracy (88.25\% vs 84.37\%) and a 6.98 percentage point improvement in F1 score (48.23\% vs 41.25\%). Precision also increases from 30.91\% to 38.16\%, indicating fewer false positives. Although ReVeal achieves slightly lower recall (60.91\% vs 66.09\%), this gain does not offset its lower precision. In security-sensitive contexts, where excessive alerts can overwhelm developers, a more balanced trade-off is preferable.

Finally, the proposed framework demonstrates the lowest standard deviation across all metrics, suggesting higher robustness to random initialisation and sampling variation. This consistency further supports its practical suitability for integration into secure development pipelines.

Overall, the results indicate that ExplainVulD provides more accurate, stable, and actionable predictions than both traditional static analysis tools and existing learning-based methods.

\subsection{Explainability of Model Predictions (RQ5)}

\begin{figure}[!t]
  \centering
  \begin{framed}
    \tiny
    \begin{tabular}{l}
    \verb|static void _UTF32ToUnicodeWithOffsets(UConverterToUnicodeArgs *pArgs,| \\
    \verb|                                          UErrorCode *pErrorCode) {| \\
    \verb|  UConverter *cnv = pArgs->converter;| \\
    \verb|  const char *source = pArgs->source, *limit = pArgs->sourceLimit;| \\
    \verb|  int32_t *offsets = pArgs->offsets, state = cnv->mode, offsetDelta = 0;| \\
    \verb|  while (source < limit && U_SUCCESS(*pErrorCode)) {| \\
    \verb|    switch (state) {| \\
    \verb|      case 0: b = *source;| \\
    \verb|        if (b == 0) state = 1;| \\
    \verb|        else if (b == (char)0xff) state = 5;| \\
    \verb|        else { state = 8; continue; } ++source; break;| \\
    \verb|      case 1 ... 7:| \\
    \verb|        if (*source == utf32BOM[state]) { ++state; ++source; ... }| \\
    \verb|        else { ...; state = 8; continue; } break;| \\
    \verb|      case 8: pArgs->source = source;| \\
    \verb|        (offsets ? T_UConverter_toUnicode_UTF32_BE_OFFSET_LOGIC| \\
    \verb|         : T_UConverter_toUnicode_UTF32_BE)(pArgs, pErrorCode);| \\
    \verb|        source = pArgs->source; break;| \\
    \verb|      case 9: pArgs->source = source;| \\
    \verb|        (offsets ? T_UConverter_toUnicode_UTF32_LE_OFFSET_LOGIC| \\
    \verb|         : T_UConverter_toUnicode_UTF32_LE)(pArgs, pErrorCode);| \\
    \verb|        source = pArgs->source; break;| \\
    \verb|      default: break;| \\
    \verb|    }| \\
    \verb|  }| \\
    \verb|  if (offsets && offsetDelta) {| \\
    \verb|    int32_t *end = pArgs->offsets; while (offsets < end) *offsets++ += offsetDelta;| \\
    \verb|  }| \\
    \verb|  pArgs->source = source;| \\
    \verb|  if (source == limit && pArgs->flush) switch (state) {| \\
    \verb|    case 8: T_UConverter_toUnicode_UTF32_BE(pArgs, pErrorCode); break;| \\
    \verb|    case 9: T_UConverter_toUnicode_UTF32_LE(pArgs, pErrorCode); break;| \\
    \verb|    default: ...; break;| \\
    \verb|  }| \\
    \verb|  cnv->mode = state;| \\
    \verb|}| 
    \end{tabular}
  \end{framed}
  \caption{Condensed version of \texttt{\_UTF32ToUnicodeWithOffsets}, simplified for readability.}
  \label{fig:utf32-condensed}
\end{figure}

\begin{figure}[!t]
\centering

\label{fig:explainability-wide}

\tiny
\begin{tabular}{|l|l|}
\hline
\multicolumn{2}{|c|}{\textbf{Prediction Summary}} \\
\hline
Predicted Class & Vulnerable (Class 1) \\
\hline
\end{tabular}

\vspace{0.5em}

\begin{tabular}{|l|l|}
\hline
\multicolumn{2}{|c|}{\textbf{Top Contributing Nodes}} \\
\hline
(505) & \texttt{ParameterType} — \texttt{UConverterToUnicodeArgs *} \\
(502) & \texttt{Identifier} — \texttt{\_UTF32ToUnicodeWithOffsets} \\
(7)   & \texttt{IdentifierDeclType} — \texttt{UConverter *} \\
(508) & \texttt{ParameterType} — \texttt{UErrorCode *} \\
(501) & \texttt{ReturnType} — \texttt{static void} \\
\hline
\end{tabular}

\vspace{0.5em}

\begin{tabular}{|l|p{6.0cm}|}
\hline
\multicolumn{2}{|c|}{\textbf{Top Contributing Edges}} \\
\hline
(769), USE &
Source: (458) \texttt{AssignmentExpression} — \texttt{pArgs → sourceLimit = pArgs → source + (state \& 3)} \\
& Target: (523) \texttt{Symbol} — \texttt{*pArgs} \\

(765), IS\_AST\_PARENT &
Source: (458) \texttt{AssignmentExpression} — \texttt{pArgs → sourceLimit = pArgs → source + (state \& 3)} \\
& Target: (29) \texttt{PtrMemberAccess} — \texttt{pArgs → sourceLimit} \\

(333), REACHES &
Source: (113) \texttt{ExpressionStatement} — \texttt{state = 8} \\
& Target: (228) \texttt{ExpressionStatement} — \texttt{pArgs → source = utf32BOM + (state \& 4)} \\

(334), REACHES &
Source: (113) \texttt{ExpressionStatement} — \texttt{state = 8} \\
& Target: (210) \texttt{Condition} — \texttt{count == (state \& 3)} \\

(337), REACHES &
Source: (113) \texttt{ExpressionStatement} — \texttt{state = 8} \\
& Target: (148) \texttt{Condition} — \texttt{state == 4} \\
\hline

\end{tabular}
\vspace{1mm}

\caption{Interpretability output from the proposed ExplainVulD. The top 5 nodes and edges are shown for a function predicted as vulnerable. Node and edge IDs refer to raw indices in the original Code Property Graph (CPG).}
\label{fig:explanation}
\end{figure}

To evaluate the explainability of our framework, we conducted a case study using a real-world function from the Chrome codebase: \texttt{\_UTF32ToUnicodeWithOffsets} (Figure~\ref{fig:utf32-condensed}). This function involves low-level control flow, pointer dereferencing, and bitwise operations, features commonly associated with memory-safety vulnerabilities. Such patterns are difficult to detect using signature-based approaches and require reasoning over both syntax and data dependencies.

The model correctly classified the function as vulnerable. To understand the basis of this decision, we examined the output of the explanation module. This component identifies the most influential nodes and edges in the corresponding Code Property Graph by combining attention weights from the GATv2 layers with relevance scores derived from input gradients. The resulting importance scores reflect both structural influence and sensitivity to the model's output. Figure~\ref{fig:explanation} shows the top five contributing nodes and edges. The most important nodes include the return type, function identifier, and parameter types. These components define the function's interface and help the model reason about input handling and memory access. Their high relevance suggests that ExplainVulD relies on semantic cues related to function contracts when assessing safety. The top-ranked edges reveal how information flows through the program. Several \texttt{REACHES} edges link the assignment \texttt{state = 8} to downstream conditional checks, such as \texttt{state == 4} and \texttt{count == (state \& 3)}. This reflects the presence of a state machine, a common construct in parsers and converters that may lead to security issues if mishandled. One \texttt{USE} edge connects the assignment \texttt{pArgs → sourceLimit = pArgs → source + (state \& 3)} to the pointer dereference \texttt{*pArgs}, highlighting data dependencies that may influence buffer access or boundary checks. These explanation outputs are consistent with how a human reviewer might reason about the function. The module isolates critical control and data flow constructs that contribute to the vulnerability prediction, providing meaningful context for developers or analysts. By mapping attention and gradient signals back to the original CPG, the framework supports traceable, post hoc explanation of each classification outcome. This case study demonstrates that ExplainVulD produces explainable predictions by identifying specific code regions that influence its decisions. The highlighted nodes and edges correspond to relevant program elements, showing that the model does not rely on spurious correlations. These qualities make the framework suitable for integration into secure development workflows, where transparency and trust in automated predictions are essential.

\section{Discussion}
\label{discussion}

This work aimed to develop a reliable and explainable framework for detecting vulnerabilities in C/C++ code, with a particular focus on handling class imbalance.

To evaluate reliability (RQ1), the model was trained and tested across 30 independent runs using randomly sampled data splits. With class-weighted cross-entropy, ExplainVulD achieved a mean accuracy of 88.25\%, F1 score of 48.23\%, and AUC of 87.64\%. Standard deviations across these metrics were low, indicating stable performance under variation in data partitioning and initialisation. These results suggest that the model generalises consistently and is robust to dataset sampling noise.

To examine how the model learns under imbalance (RQ2), we explored several loss functions during development. Although focal loss configurations yielded improvements in precision, class-weighted cross-entropy provided the most balanced trade-off between precision, recall, and stability. This outcome demonstrates the practicality of weighted loss functions in real-world imbalanced settings.

We then investigated the contribution of input features (RQ3). Semantic features alone resulted in poor generalisation, with predictions dominated by the majority class. Structural features capturing control and data flow improved performance. Combining both semantic and structural embeddings yielded further gains, and the addition of edge-type encodings produced the highest overall results. These findings highlight the importance of representing both code content and structure, along with the relationships between program elements.

To assess comparative effectiveness (RQ4), we benchmarked ExplainVulD against ReVeal and two static analysis tools. The framework achieved higher accuracy, precision, recall and F1 score than all baselines including two static analysers and ReVeal model. The lower variance in ExplainVulD's results further supports its reliability across runs.

Explainability was evaluated through a case study using a real-world vulnerable function from the Chrome codebase (RQ5). The model correctly identified the function as vulnerable, and the explanation module highlighted nodes and edges in the Code Property Graph that corresponded to key program elements, including parameter types, and data and control dependencies. These explanations aligned with human-understandable reasoning, suggesting that the model’s predictions are grounded in meaningful program structure. This supports the use of ExplainVulD in secure development workflows that require transparency as well as predictive accuracy.

\subsection{Limitations and Future Work}

While ExplainVulD achieves consistent results under imbalanced conditions, there are some limitations to consider. The evaluation is based on the ReVeal dataset, which includes functions from two large-scale codebases: Chrome and Debian. Although these are realistic and security-relevant, broader validation across additional projects would be needed to assess generalisability more fully. The framework also operates on static Code Property Graphs, which may limit its ability to capture dynamic behaviours such as runtime control flow or heap state. Finally, the dual-channel embedding design and edge-aware attention mechanism introduce additional model complexity, which may affect training efficiency in low-resource settings.

These aspects motivate several directions for future work. One is to extend the framework to multiclass vulnerability classification, enabling finer-grained identification of flaw types such as buffer overflows or use-after-free errors. Another is to explore cross-project evaluation, training on one codebase and testing on another, to better reflect deployment conditions. Incorporating dynamic signals, such as execution traces or inter-procedural control flow, could further improve detection in cases where static structure is insufficient.

\section{Conclusion}
\label{conclusion}

This paper presented ExplainVulD, a graph-based framework for vulnerability detection in C/C++ code. The approach integrates semantic and structural information through dual-channel node embeddings and applies an edge-aware variant of GATv2 to reason over heterogeneous code graphs. The framework was evaluated using 30 independent runs on the ReVeal dataset, demonstrating consistent performance in terms of accuracy, F1 score, and AUC under severe class imbalance. Ablation studies showed that each component, including semantic features, structural patterns, and edge-type encodings, contributes to performance, with their combination yielding the most effective results. In comparison with both learning-based baselines and traditional static analysis tools, ExplainVulD achieved a more favourable balance between false positives and false negatives. A case study further illustrated that the explanation module highlights relevant subgraphs, aligning with source-level vulnerability reasoning. While the results are promising, this study is limited to the ReVeal dataset, which includes two codebases. Further evaluation on a broader range of projects is required to assess generalisation. The use of static graphs also constrains visibility into dynamic behaviours such as heap state and runtime control flow, which are important for certain vulnerability types. These aspects motivate future work on incorporating dynamic analysis and evaluating cross-project performance. ExplainVulD contributes a practical and explainable approach to graph-based vulnerability detection. Its design supports integration into secure software development workflows, offering early-stage insights that may assist developers in identifying and understanding potential security flaws.







\begin{thebibliography}{10}

\bibitem{max-mergin}
V.~Nguyen, T.~Le, C.~Tantithamthavorn, J.~Grundy, H.~Nguyen, and D.~Phung, ``Cross project software vulnerability detection via domain adaptation and max-margin principle,'' 2022.

\bibitem{wheeler_flawfinder}
D.~A. Wheeler, ``Flawfinder.'' \url{https://dwheeler.com/flawfinder/}, 2025.
\newblock Accessed: 2025-06-10.

\bibitem{cppcheck}
{Cppcheck Project}, ``Cppcheck: A tool for static c/c++ code analysis.'' \url{https://cppcheck.sourceforge.io/}, 2024.
\newblock Accessed: 2025-06-18.

\bibitem{1quDeASK}
J.~Smith, B.~Johnson, E.~Murphy-Hill, B.~Chu, and H.~R. Lipford, ``Questions developers ask while diagnosing potential security vulnerabilities with static analysis,'' ESEC/FSE 2015, (New York, NY, USA), p.~248–259, Association for Computing Machinery, 2015.

\bibitem{whynotstatic}
B.~Johnson, Y.~Song, E.~Murphy-Hill, and R.~Bowdidge, ``Why don't software developers use static analysis tools to find bugs?,'' in {\em 2013 35th International Conference on Software Engineering (ICSE)}, pp.~672--681, 2013.

\bibitem{zou2019uvuldeepecker}
D.~Zou, Y.~Wang, X.~Ou, H.~Jin, and Z.~Deng, ``\ensuremath{\mu}vuldeepecker: A deep learning-based system for multiclass vulnerability detection,'' {\em IEEE Transactions on Dependable and Secure Computing}, 2019.

\bibitem{zou2019mu}
D.~Zou, S.~Wang, S.~Xu, Z.~Li, and H.~Jin, ``$\mu$vuldeepecker: A deep learning-based system for multiclass vulnerability detection,'' {\em IEEE Transactions on Dependable and Secure Computing}, vol.~18, no.~5, pp.~2224--2236, 2019.

\bibitem{9321538-sysevr}
Z.~Li, D.~Zou, S.~Xu, H.~Jin, Y.~Zhu, and Z.~Chen, ``Sysevr: A framework for using deep learning to detect software vulnerabilities,'' {\em IEEE Transactions on Dependable and Secure Computing}, vol.~19, no.~4, pp.~2244--2258, 2022.

\bibitem{vuldelocator}
Z.~Li, D.~Zou, S.~Xu, Z.~Chen, Y.~Zhu, and H.~Jin, ``Vuldeelocator: A deep learning-based fine-grained vulnerability detector,'' {\em IEEE Transactions on Dependable and Secure Computing}, vol.~19, p.~2821–2837, July 2022.

\bibitem{zhou2019devign}
Y.~Zhou, S.~Liu, J.~Siow, X.~Du, and Y.~Liu, ``Devign: Effective vulnerability identification by learning comprehensive program semantics via graph neural networks,'' {\em Advances in neural information processing systems}, vol.~32, 2019.

\bibitem{BGNN4VD}
S.~Cao, X.~Sun, L.~Bo, Y.~Wei, and B.~Li, ``Bgnn4vd: Constructing bidirectional graph neural-network for vulnerability detection,'' {\em Information and Software Technology}, vol.~136, p.~106576, 2021.

\bibitem{chakraborty2021deep}
S.~Chakraborty, R.~Krishna, Y.~Ding, and B.~Ray, ``Deep learning based vulnerability detection: Are we there yet?,'' {\em IEEE Transactions on Software Engineering}, vol.~48, no.~9, pp.~3280--3296, 2021.

\bibitem{1ivdetc}
Y.~Li, S.~Wang, and T.~N. Nguyen, ``Vulnerability detection with fine-grained interpretations,'' in {\em Proceedings of the 29th ACM Joint Meeting on European Software Engineering Conference and Symposium on the Foundations of Software Engineering}, ESEC/FSE 2021, (New York, NY, USA), p.~292–303, Association for Computing Machinery, 2021.

\bibitem{cpg}
F.~Yamaguchi, N.~Golde, D.~Arp, and K.~Rieck, ``Modeling and discovering vulnerabilities with code property graphs,'' in {\em 2014 IEEE Symposium on Security and Privacy}, pp.~590--604, 2014.

\bibitem{mikolov2013efficientestimationwordrepresentations}
T.~Mikolov, K.~Chen, G.~Corrado, and J.~Dean, ``Efficient estimation of word representations in vector space,'' 2013.

\bibitem{gatv2}
S.~Brody, U.~Alon, and E.~Yahav, ``How attentive are graph attention networks?,'' {\em arXiv preprint arXiv:2105.14491}, 2021.

\bibitem{ying2019gnnexplainergeneratingexplanationsgraph}
R.~Ying, D.~Bourgeois, J.~You, M.~Zitnik, and J.~Leskovec, ``Gnnexplainer: Generating explanations for graph neural networks,'' 2019.

\bibitem{KAUR20202023-4}
A.~Kaur and R.~Nayyar, ``A comparative study of static code analysis tools for vulnerability detection in c/c++ and java source code,'' {\em Procedia Computer Science}, vol.~171, pp.~2023--2029, 2020.
\newblock Third International Conference on Computing and Network Communications (CoCoNet'19).

\bibitem{russell2018automated}
R.~Russell, L.~Kim, L.~Hamilton, T.~Lazovich, J.~Harer, O.~Ozdemir, P.~Ellingwood, and M.~McConley, ``Automated vulnerability detection in source code using deep representation learning,'' in {\em 2018 17th IEEE international conference on machine learning and applications (ICMLA)}, pp.~757--762, IEEE, 2018.

\bibitem{8599360-19}
A.~Xu, T.~Dai, H.~Chen, Z.~Ming, and W.~Li, ``Vulnerability detection for source code using contextual lstm,'' in {\em 2018 5th International Conference on Systems and Informatics (ICSAI)}, pp.~1225--1230, 2018.

\bibitem{https://doi.org/10.1155/2021/5566423-kelm}
G.~Tang, L.~Yang, S.~Ren, L.~Meng, F.~Yang, and H.~Wang, ``An automatic source code vulnerability detection approach based on kelm,'' {\em Security and Communication Networks}, vol.~2021, no.~1, p.~5566423, 2021.

\bibitem{fine-tune_llm}
A.~Shestov, R.~Levichev, R.~Mussabayev, E.~Maslov, P.~Zadorozhny, A.~Cheshkov, R.~Mussabayev, A.~Toleu, G.~Tolegen, and A.~Krassovitskiy, ``Finetuning large language models for vulnerability detection,'' {\em IEEE Access}, vol.~13, pp.~38889--38900, 2025.

\bibitem{TANG2023111623-csgvd}
W.~Tang, M.~Tang, M.~Ban, Z.~Zhao, and M.~Feng, ``Csgvd: A deep learning approach combining sequence and graph embedding for source code vulnerability detection,'' {\em Journal of Systems and Software}, vol.~199, p.~111623, 2023.

\bibitem{10.1145/3510454.3516865-regvd}
V.-A. Nguyen, D.~Q. Nguyen, V.~Nguyen, T.~Le, Q.~H. Tran, and D.~Phung, ``Regvd: revisiting graph neural networks for vulnerability detection,'' ICSE '22, (New York, NY, USA), p.~178–182, Association for Computing Machinery, 2022.

\bibitem{WARTSCHINSKI2022106809-vudenc}
L.~Wartschinski, Y.~Noller, T.~Vogel, T.~Kehrer, and L.~Grunske, ``Vudenc: Vulnerability detection with deep learning on a natural codebase for python,'' {\em Information and Software Technology}, vol.~144, p.~106809, 2022.

\bibitem{code2vec}
P.~Bojanowski, E.~Grave, A.~Joulin, and T.~Mikolov, ``Enriching word vectors with subword information,'' {\em Transactions of the association for computational linguistics}, vol.~5, pp.~135--146, 2017.

\bibitem{MLP}
F.~Murtagh, ``Multilayer perceptrons for classification and regression,'' {\em Neurocomputing}, vol.~2, no.~5, pp.~183--197, 1991.

\bibitem{black2017sard}
P.~Black, ``{SARD: Thousands of Reference Programs for Software Assurance},'' {\em Journal of Cyber Security and Information Systems}, 2017.
\newblock \url{https://tsapps.nist.gov/publication/get_pdf.cfm?pub_id=923127} (Accessed June 11, 2025).

\bibitem{10.5555/249049}
M.~T. Hagan, H.~B. Demuth, and M.~Beale, {\em Neural network design}.
\newblock USA: PWS Publishing Co., 1997.

\bibitem{wheeler2025flawfinder}
D.~A. Wheeler, ``{Flawfinder}: A static analysis tool for finding vulnerabilities in c/c++.'' \url{https://dwheeler.com/flawfinder/}, 2025.
\newblock Accessed June 17, 2025.

\bibitem{infer2025}
``{Infer}: A static analysis tool for java, c/c++, and objective-c.'' \url{https://fbinfer.com/}, 2025.
\newblock Accessed June 17, 2025. Developed and maintained by Meta (formerly Facebook).

\bibitem{focal}
T.-Y. Lin, P.~Goyal, R.~Girshick, K.~He, and P.~Dollár, ``Focal loss for dense object detection,'' 2018.

\end{thebibliography}
\end{document}